\documentclass[a4paper,fleqn]{cas-dc}

\usepackage[numbers]{natbib}
\usepackage{graphicx}
\usepackage{multirow}
\usepackage{booktabs}
\usepackage{amsmath}
\usepackage{makecell} %
\usepackage{url} %

\newcommand{\sysname}{BARBICAN}

\begin{document}
\let\WriteBookmarks\relax
\def\floatpagepagefraction{1}
\def\textpagefraction{.001}

\shorttitle{Component Attribution in \sysname}
\shortauthors{İ. Abasıkeleş-Turgut and E. Gümüş}

\title[mode=title]{Deletion, Substitution, Repair and Transplant: Component Attribution in \sysname, a Layered Prompt-Injection Gateway for MCP}

\author[1]{{\.I}pek Abasıkeleş-Turgut}[orcid=0000-0002-5068-969X]
\cormark[1]
\ead{ipek.abasikeles@iste.edu.tr}
\credit{Conceptualization, Methodology, Formal analysis, Investigation,
Validation, Writing -- original draft, Writing -- review \& editing, Supervision}

\affiliation[1]{organization={Department of Computer Engineering, Faculty of Engineering and Natural Sciences},
                addressline={Iskenderun Technical University, Merkez Kampüs},
                city={İskenderun},
                postcode={31200},
                state={Hatay},
                country={Türkiye}}

\author[2]{Edip Gümüş}[orcid=0009-0009-3000-7645]
\ead{edipgumus.lee25@iste.edu.tr}
\credit{Software, Data curation, Investigation, Resources, Writing -- review \&
editing}

\affiliation[2]{organization={Department of Computer Engineering, Institute of Graduate Studies},
                addressline={Iskenderun Technical University, Merkez Kampüs},
                city={İskenderun},
                postcode={31200},
                state={Hatay},
                country={Türkiye}}

\cortext[1]{Corresponding author}

\begin{abstract}
A neural classifier can fail completely inside a layered defense without any
end-to-end number moving. We demonstrate this on \sysname, a fail-closed
gateway for language-model tool mediation whose arbitration stage spent an entire recorded
evaluation receiving text mangled by a misconfigured tokenizer. Correcting the
misconfiguration lifts that stage's corpus-wide area under the ROC curve (AUC) from $0.392$ to $0.855$ and
pushes system recall down, from $80.67\%$ to $80.34\%$; a stub that answers
``malicious'' scores $80.77\%$, higher than either. The reason emerges once the
stage's input distribution is examined: upstream routing hands it a band that is
$96.82\%$ malicious, and the enforcement design lets an incriminating verdict
block while an exculpating one releases an input only through one narrow
branch. Yet deleting it costs 673
detections. Substituting
its verdict with a constant instead moves between 12 and 140 of 25,000 decisions.
Transplanted into the arbitration stage, an off-the-shelf classifier that
separates the band's inputs at an AUC of $0.916$ changes 16 of 25,000 decisions.
Measured against such detectors with operating points chosen away from the test
data, the gateway leads on its own benchmark distribution and falls behind once
that distribution changes. In our earlier gateway, the review model's band is $92.94\%$
malicious and a constant verdict is as accurate as the review model. We argue
that an ablation figure for any routed stage is unreadable without the class
balance of the band that stage receives, and that deletion experiments need
substitution experiments beside them.
\end{abstract}

\begin{highlights}
\item A classifier repaired from AUC $0.392$ to $0.855$ makes its host system detect less
\item Under the gateway's own enforcement, deletion and substitution differ 4.8-fold
\item Upstream routing leaves the arbitration band $96.82\%$ malicious
\item Benchmark labels encode prompt form, not harm; a public classifier leads under shift
\item A stronger classifier transplanted into the arbitration stage moves 16 decisions
\end{highlights}


\maketitle

\section{Introduction}\label{sec:intro}

Language models increasingly act rather than only answer, and interfaces such as
the Model Context Protocol (MCP) have made that routine. The exposure this creates is
structural: whatever a tool returns enters the model's context as text, and text
in context is instruction-shaped. A retrieved document, a code comment or an
error message can carry directives that the model has no reliable basis for
distinguishing from its principal's own. Defenses against this have converged on
a recognizable shape---deterministic rules dispose of the obvious cases, a
semantic layer catches paraphrase, an expensive neural or LLM arbitrator
adjudicates what survives, and a fail-closed default blocks whatever remains
unresolved.

Each of those stages costs latency, memory or an inference call, so a layered
defense must eventually say which of them earns its place. The standard
instrument is ablation by deletion: switch a layer off, re-run the benchmark,
attribute the drop in detection to the layer removed---an experiment one recent
gateway describes as a component ablation that ``isolates each layer's
contribution'' \cite{spotlightguard2026}. Isolation is the claim worth examining. In a cascade
a stage is not an independent module but a position, defined by the routing rule
that decides what reaches it and by the policy that disposes of what it leaves
unresolved; switching the stage off removes all three at once, and the resulting
drop is their joint contribution, reported as one component's. Outside security
the point is settled---zero-ablation can overstate a neural component's
importance several-fold, because it moves the network off the data manifold
rather than removing a function \cite{li2024}, and statisticians made the
corresponding argument about dropped covariates earlier
\cite{hooker2021,lei2018}---yet deletion remains the convention by which layered
defenses report what their layers contribute.

What that convention costs in practice cannot be settled from the convention
itself. It needs a system that can be instrumented rather than inspected: one
whose source, configuration history and audit trail are available, and whose
evaluation can be re-executed rather than re-read. \sysname{} is such a
system---a fail-closed gateway mediating MCP tool calls, developed within our own
research group and renamed here so that it and its artifacts can be cited under a
single stable identifier. Its recorded evaluation, $88.26\%$ accuracy over a
balanced benchmark of 25,000 prompts, is logged in the artifact manifest of the
code revision we re-execute, and our control run reproduces its confusion matrix
sample for sample \cite{barbican2026}; every measurement below is a departure
from it. We regard
our authorship of the system as a condition of the study rather than as a caveat:
the interventions that follow require the source, the configuration history and
the audit trail, together with a willingness to publish what they turn up, and an
outside evaluator would have had none of the first three and every incentive to
avoid the fourth.

That baseline conceals a component failure. Throughout the run that produced it,
the gateway's neural arbitration stage was reading text that a misconfigured
tokenizer had filled with unknown-token markers, on a model that classifies its
own validation set perfectly. Measured on its own, the stage separated malicious
from benign inputs at an AUC of $0.392$---worse than chance, and informative only
when read in reverse. None of the
accuracy, precision, recall, F1 or correlation figures the gateway recorded gave
any sign of it.

Correcting the tokenizer does not restore a lost result; it makes matters
stranger. The repaired stage reaches an AUC of $0.855$, and the system's recall
falls, from $80.67\%$ to $80.34\%$: repairing the classifier costs 41 detections
net, 48 lost and 7 gained.
Replacing it with three lines of code that answer ``malicious'' without examining
their input raises recall instead, to $80.77\%$. On the four public corpora
inside the benchmark, correcting the classifier moves recall by $0.00$ percentage
points. Switching the stage off entirely costs 673 detections---the figure an
ablation table would print, and the figure a reader would take as the stage's
worth.

These numbers cannot be reconciled as sampling noise, and their spread is this
paper's point of entry. Pinning the stage's verdict to a constant, with routing,
thresholds and enforcement left intact, moves between 12 and 140 of 25,000
decisions; the two sets do not overlap, so the stage's real influence is bounded
by 152 decisions, $0.61\%$. Deleting the
stage, which in this gateway means letting its band fail open, moves 673. The
4.8-fold discrepancy is not a measurement artifact but a consequence of what each
operation removes: deletion takes the band out of review, so that the fail-closed
rule covering unresolved reviews no longer applies and a looser release rule for
allowed inputs does, and the detections lost belong mostly to that change.
Both operations lose their detections through releases with weak residual
evidence, so the size of the discrepancy depends on how those releases are
enforced (Section~\ref{sec:results-mapping}).

The mechanism becomes legible once the stage's input distribution is examined
rather than assumed. Cascades are built so that expensive stages see little
traffic, and the traffic they do see is not a random sample: it is what the
earlier layers could neither clear nor condemn. In \sysname{} that band is
$96.82\%$ malicious, against $30.11\%$ outside it. Inside a population that
skewed, refusing to discriminate is a strong policy---a constant positive earns
$96.82\%$ precision---and discriminating well is expensive, because almost every
input the classifier declines to condemn deserved condemning. The enforcement
path compounds this asymmetrically: a verdict of ``malicious'' ends in a block,
by escalation or by the fail-closed rule, while a verdict of ``benign'' is checked against further
conditions before it may release anything, and with the gateway's own semantic
scores those conditions almost never pass: never in the control, 12 times in
7,453 when every verdict is forced benign. With those scores the stage's output is
close to a one-way signal, able to add blocks and nearly unable to remove them.
Under these two conditions an arbitrator can gain little over a constant,
whatever its quality, and accuracy converts into lost coverage rather than into
avoided false positives; the ceiling limits what it can gain, not what it can
lose (Section~\ref{sec:results-transplant}).

If the arbitration stage is not producing the reported detections, something else
is. Tracing them to their sources places them in an undocumented fusion stage, in
a set of hand-written policy branches keyed on formatting strings, four of
the five firing only on synthetic samples and never on public corpora, in a
similarity memory whose added blocks are correct $55.6\%$ of the time, and in the
fail-closed conversion itself. None of these appears as a component in the
system's own architecture description.

The evidence is fifteen live re-executions of \sysname{} over its own
25,000-sample benchmark, with the corpus and the machine held fixed, the code
pinned to one of two recorded states (Section~\ref{sec:protocol}), and three
operations applied to the arbitration stage: deleting it, which
reproduces what an ablation study would report; pinning its verdict to a
constant, which isolates what its decisions are worth; and correcting its
tokenizer, which reveals whether the system responds to the stage's quality at
all. A control configuration with every switch unset reproduces the recorded
confusion matrix without a single differing sample, and that agreement is what
licenses reading the remaining fourteen runs as departures from it.
A second series of twelve runs repeats the control three times, isolates the
post-allow escalation step that the first series left unmeasured, and sweeps the
semantic threshold; and an independent re-execution on different hardware
reproduces the control's confusion matrix exactly, with two of 25,000 decisions
exchanged.

Two further questions are answered by measurement rather than by caveat. The
first concerns the corpus. Tracing every public record back to its source shows
that one public collection is labeled by the form of a prompt rather than by
its harm, that a second has had its attack strings embedded in generated HTML
that the original benchmark does not contain, and that a third carries an
appended line of challenge metadata; the gateway's per-source figures have to be
read through these facts. The second concerns comparison. Four public detectors
were run over the same records, with every operating point chosen on data the
reported figure does not use. On the gateway's own distribution it leads them;
where the distribution changes, the strongest of them dominates it on detection
and on false alarms together.

That comparison invites a fourth operation, which we call transplant: leave the
gateway's routing, thresholds and enforcement untouched and put a stronger
classifier in one of its stages. Repair asks whether the system tracks the
quality of its own component; transplant asks the same of a component known to be
good. Placed in the arbitration stage, a public classifier that separates the
band's inputs at an AUC of $0.916$, against $0.368$ for the gateway's own
classifier on the same band, changes 16 of 25,000 decisions. It labels 202 of the
band's benign inputs benign, and none of them is released; the four inputs it
releases are attacks. Placed in the semantic
stage instead, the same classifier changes 1,015 decisions, because there it
decides what is routed rather than what is said about it.

From these measurements we draw a reporting proposal. Any stage receiving routed
rather than complete traffic should be reported together with the class balance
of the band routed to it, because that balance caps what the stage can contribute
before its quality is considered at all; and because deletion and substitution
answer different questions, a table reporting one should report both. Neither
requirement costs more than an additional run.

Two boundaries should be stated at the outset. This is not a comparison of
complete defenses: no competing gateway was rebuilt and re-measured under this
protocol, and the four detectors we run beside \sysname{} are single classifiers
used as external baselines, so no claim is made about how \sysname{} ranks
against any other gateway. %
Two other layered defenses, LLM Guard and NeMo Guardrails, are run on the
independent check set component by component, and as complete stacks on the
corpus or a sample of it (Section~\ref{sec:results-others}),
to see what deletion credits in stacks that do not route; that is an audit of
their attributions, not a ranking. Neither is it a replication
in the usual sense, since the recorded confusion matrix reproduces exactly; the
disagreement concerns what that result attributes to which component, not whether
it holds.

Section~\ref{sec:related} places the deletion convention in the layered-defense
literature and the intervention alternatives in the interpretability and
statistics literatures. Section~\ref{sec:system} describes \sysname{} twice, as
documented and as instrumented. Section~\ref{sec:method} defines the four
operations and the band base rate and states the protocol.
Section~\ref{sec:results} reports the fifteen runs, the second series, the
corpus audit, the external comparison, an independent check set, the
transplants, two further defenses, our earlier gateway and a second platform. Section~\ref{sec:missed} asks
why no recorded metric exposed either of the two faults, and lists what would
have. Sections~\ref{sec:limitations} and~\ref{sec:conclusion} close.

\section{Related Work}\label{sec:related}

\subsection{Layered defenses, and how they credit their layers}

The Model Context Protocol gives language-model applications a uniform way to
discover and call external tools, and with it a path from untrusted text to file
access, database queries and network requests
\cite{mcpspec,hou2025},
a surface whose threat taxonomy is now being standardized \cite{owaspmcp}.
Defenses published since have converged on a recognizable shape: a deterministic
pre-filter, a semantic stage, a small classifier or local language model that
arbitrates the remainder, and a policy layer mapping the result onto allow,
review and block. ShieldMCP intercepts at the tool layer and combines structural
analysis with semantic intent validation \cite{shieldmcp2026}; a threat-modeling
study of tool poisoning recommends the same multi-layer arrangement
\cite{huang2026}; and dependency-graph and benchmark work approaches the surface
from other directions \cite{mcptox,mindguard,jamshidi2025}.
\sysname{} belongs to this family, and this paper is not a proposal to replace
it. What we examine is how systems of this shape are evaluated.

The attack itself was first described in its indirect form, in which
instructions planted in retrieved content reach the model at inference time
\cite{greshake2023}; it has since been formalized and benchmarked against
defenses \cite{liu2024formalizing} and evaluated in dynamic agent environments
where tools are called over untrusted data \cite{agentdojo2024}. A second family
of defenses acts on the model or on the surrounding system rather than on the
input stream: marking the provenance of untrusted text inside the prompt
\cite{spotlighting2024}, separating prompt and data into two channels and
fine-tuning the model to follow only the first \cite{struq2025}, preference
optimization toward the legitimate instruction \cite{secalign2025}, and a
protective system layer that keeps untrusted data from changing the program's
control flow \cite{camel2026}. A recent replication of several defenses in such
an environment argues that their evaluations should report legitimate-task
utility and run-to-run variation alongside attack success \cite{khan2026}. The
systems studied here belong to the layered family described above.

On that question the field is remarkably uniform. A component is disabled, the
benchmark re-run, and the drop in a headline metric read as that component's
contribution. Spotlight-Guard reports that ``a component ablation isolates each
layer's contribution'' \cite{spotlightguard2026}; PromptGuard
quantifies the ``individual and combined contributions'' of its
regular-expression and MiniBERT components \cite{promptguard2026}; and the
benchmark-and-defense study accompanying BIPIA gauges each component's
contribution to the overall defense \cite{yi2025}. The
convention is not peculiar to language-model security either: work on intrusion
detection for vehicular networks reports that ``an ablation study was conducted to
isolate the contribution of each of its modules'' \cite{evhybrid2025}.

Our own earlier work is part of this pattern, which is why we do not present the
convention as someone else's habit. In describing a three-layer MCP gateway we
reported nested configurations---rule layer alone, rule layer with a semantic
stage, both with a local review model---and read the differences between them as
component contributions \cite{cascade}. That paper did observe that its review
stage altered no final decision, which is a substitution-style result arrived at
incidentally; what it did not do was separate the two kinds of measurement, and
the present paper is in part a correction of that omission. Neither there nor,
as far as we can establish, in any of the layered defenses above is the class balance of the
traffic reaching an ablated stage reported. Section~\ref{sec:baserate} argues
that without that number the ablation figures are not interpretable.
Section~\ref{sec:results-cascade} computes it for that earlier gateway.

\subsection{Deletion, intervention, and variable importance}

That removing a part overstates its importance is an established result outside
security, reached independently in two literatures. In mechanistic
interpretability, Li and Janson show that zero-, mean- and resample-ablation
systematically exaggerate component importance in transformer circuits, the
optimal-ablation estimate falling to a small fraction of the zero-ablation
estimate on standard tasks \cite{li2024}, and Leavitt and Morcos argue directly for
intervention over correlational measures of component relevance
\cite{leavitt2021}. Activation patching and interchange interventions, together
with the causal-abstraction framework that unifies them, supply the formal
vocabulary for substituting a component's output rather than deleting the
component \cite{zhangnanda2024,geiger2025,vig2020,geiger2021};
measuring importance by removal is by now automated
\cite{conmy2023,meng2022,wang2023},
and its interpretation actively debated \cite{makelov2024}.

In statistics the objection is older still. Permuting or dropping a covariate is
held not to measure its importance without at least one further model
\cite{hooker2021}; leave-one-covariate-out has a formal treatment and known
limits \cite{lei2018};
and conditional and Shapley-based alternatives were developed precisely because
marginal removal misleads under dependence
\cite{strobl2008,williamson2023,verdinelli2024}.
The common structure of both objections is that removal changes the input
distribution the remaining system faces, so that what is measured is a different
system rather than the same system without one part. A layered defense is exactly
such a case, since each stage exists to receive what the previous stage routed to
it. Yet none of this literature addresses layered security systems, and none of
the security work cited above cites it. Carrying the distinction across is one
contribution of this paper; measuring it on a system that can be re-executed is
the other.

\subsection{When end-to-end performance stops tracking component quality}

A second line of work explains why a better component need not yield a better
system. Sculley et al.\ describe entangled machine-learning systems in which
changing anything changes everything, and note that improving a single
component's model can worsen system accuracy once the residual errors become more
correlated \cite{sculley2015}, although the observation is made in an ensemble
setting rather than a cascade. For cascades specifically, Saberian and
Vasconcelos formalize why optimizing each stage against its own local objective
does not optimize the cascade as a whole \cite{saberian2014}; recent work on
compound systems assumes that end-to-end performance tracks module performance,
and work on cascades states when deferral by confidence suffices, conditions a
deployed pipeline need not satisfy \cite{chen2025selection,jitkrittum2023};
and stage-wise ablation in cascaded detectors is likewise non-monotone in
practice \cite{cairccnn2019}.

The security literature frames the same phenomenon in operational terms.
Axelsson's base-rate analysis explains why a detector's practical value is
governed by the prevalence of what it looks for \cite{axelsson2000}, and Sommer
and Paxson identify the gap between a detection output and its operational
interpretation as a standing difficulty for machine learning in intrusion
detection \cite{sommer2010}. Section~\ref{sec:baserate} applies the first of
these in the opposite direction from its usual use: routing does not lower the
base rate seen by the arbitration stage, it raises it, and that is what renders
the stage's accuracy inconsequential.

\subsection{Evaluation validity, and re-executing a single system}

Whether an ablation figure can be trusted is part of a larger question about what
security evaluations measure. Arp et al.\ enumerate the pitfalls that recur in
machine learning for computer security and quantify their prevalence across
thirty top-tier papers \cite{arp2022}; three of their categories bear on the
measurements reported here, namely sampling bias, spurious correlation, and data
snooping in its selective form, that is, tuning an operating point on the data
used to report it. Distribution shift over time and space has a parallel
treatment in malware classification \cite{pendlebury2019},
and contamination of evaluation data by training material is now surveyed in its
own right \cite{cheng2025}.
What a corpus is made of matters as much as how it is split: defenses tuned on
curated injection datasets overfit known patterns and degrade on unseen phrasing
\cite{promptsleuth},
evaluations that exclude out-of-distribution material from feature design, model
selection \emph{and} threshold tuning report markedly different numbers from
those that do not \cite{regimedep},
and in adjacent detection problems the gap between curated benchmarks and field
data is large enough to halve reported discrimination
\cite{deepfakeeval}.
Section~\ref{sec:results-corpus} reports \sysname{} separately by corpus source
for this reason.

The usual comparison points for a layered defense are standalone guardrails such
as Llama Guard and programmable rail toolkits such as NeMo Guardrails
\cite{llamaguard,nemo},
and systematic benchmarking finds that additional rails raise detection and false
positives together \cite{advprompteval}.
Taxonomies organizing guardrails by intervention stage and technique
\cite{sokjailbreak}
place the tiers of \sysname{} in a common frame, and standardized evaluation
protocols exist for jailbreak defenses specifically \cite{jailbreakbench}; two of
the corpora in the benchmark studied here come from this line
\cite{injecagent,wildteaming}. %
We make no comparative claim against complete systems of this kind; of the
standalone classifiers, Llama Guard~3 \cite{llama3herd,llamaguard3card} and three
prompt-injection classifiers \cite{promptguard2card,protectaiv2} are run here
as external baselines over the same records
(Section~\ref{sec:results-external}).

Methodologically the paper belongs instead with work that re-executes a system
and asks where its reported performance came from. Adversarial-example defenses
have been re-evaluated at scale with the conclusion that the evaluation method,
not the defense, determined the published result
\cite{athalye2018,tramer2020,carlini2019}.
In security specifically, a reproducibility study of 744 machine-learning papers
at four top venues found artifacts available for two in five and functional for
fewer than one in five \cite{olszewski2023}; a systematization of fuzzing
evaluation practice reached comparable conclusions \cite{schloegel2024}; and a
case study of one alert-classification system attributes its reported performance
largely to label imbalance in its input data \cite{teuwen2025}. The last of these
is the closest precedent for what follows, differing mainly in that the system
re-examined here is our own.

\section{The Gateway Under Study}\label{sec:system}

A component ablation can only be as informative as the architectural account it
is read against, so this section gives two accounts of \sysname{}: the one its
design documentation provides, and the one its own traces support. They differ,
and the differences are not incidental to what follows---each of them changes
what an ablation figure for this system can be taken to mean. We therefore set
the two side by side rather than reconciling them.

\subsection{As documented}\label{sec:system-documented}

\sysname{} is an inline, fail-closed gateway sitting between an LLM client and
the MCP servers it calls, and its design principle is computational tiering:
cheap, high-precision filters are expected to absorb the bulk of traffic, so that
heavier neural reasoning can be held back for borderline cases. The design
comprises an ingress layer and four tiers, closed by a safety rail.

\begin{description}
\item[Ingress middleware] Components that decode layered obfuscation, attach
provenance labels to context tokens, and validate outbound tool calls against
registered JSON schemas.

\item[L1, deterministic rules] Regular-expression signatures for known jailbreak
patterns, a control-intent density statistic, and structural signals for shell,
SQL and path-traversal constructs.

\item[L2, dense semantic layer] A multilingual sentence embedding model scores
each input against reference centroids standing for verified prompt-injection and
tool-abuse categories, fused with lexical and control-density signals into a
continuous risk score.

\item[L2.5, adaptive percentile calibrator] A rolling window over recent scores
yields empirical percentiles, from which review and block thresholds are derived.
Interquartile-range outlier rejection is described as preventing an adversary
from poisoning the window, and hard ceilings as bounding the thresholds above.

\item[L3, edge arbitrator] Inputs whose score falls between the review and block
thresholds are referred to a quantized DeBERTa-v3-small classifier running
locally under ONNX Runtime.

\item[Allow-hint safety rail] A deterministic gate permitting a benign arbitrator
verdict to de-escalate a referred input only when four further conditions hold,
so that a neural model cannot unilaterally clear a threat.
\end{description}

\subsection{As instrumented}\label{sec:system-instrumented}

Emitting per-stage state for all 25,000 prompts (Section~\ref{sec:protocol})
gives a second description, and it departs from the first in the two ways that
matter for an ablation: the named tiers do not behave as described, and they are
not the only components deciding.

Taking the tiers in turn, the semantic layer scores against 215 stored texts
rather than a set of attack centroids: 23 category references and 192 support
examples, embedded with \texttt{intfloat/multilingual-e5-small}. The calibration
stage above it does not behave adaptively either. Across all 25,000 prompts the
review threshold takes two values and the block threshold one, namely
$(0.36, 0.49)$ for 13,321 prompts and $(0.33, 0.49)$ for 11,679, with the lower
review value chosen when a risk signal is present and the higher otherwise; the
documented ceiling of $0.42$ on the review threshold is never reached (the
block threshold sits at its own ceiling, $0.49$; the semantic threshold of $0.42$
named below is a separate setting), and the rolling window never moves
the threshold off these two points.
The stage does hold one rule that reads earlier requests, and the documentation
does not mention it: once more than $62.5\%$ of a user's requests so far have not
been allowed, both thresholds drop by $0.10$. The benchmark sends every prompt as
one user, whose share of non-allowed requests ends near $42\%$ in the control
run, and the rule never fires there; across the other runs reported here it fires
only for four of the first seven prompts of one transplant run, and changes at
most one decision
(Section~\ref{sec:results-postallow} shows a run in which it fires throughout). The arbitrator itself is likewise not quite
the documented artifact, the deployed export being an INT8 ONNX model holding
141.9M parameters in total---the 44M figure in the documentation refers to the
transformer backbone excluding the embedding matrix---and its inference tokenizer
having been loaded with a compatibility flag belonging to a different model
family, which corrupts its output (Section~\ref{sec:results-repair}).

The decision surface is also wider than four tiers. Beyond them the traces expose
a fusion stage combining seven weighted signals---\texttt{combined},
\texttt{signal}, \texttt{anomaly}, \texttt{intent}, \texttt{provenance},
\texttt{l1\_obfuscation} and \texttt{l1\_control\_density}---into a single score,
and a policy stage emitting one of ten reasons. Neither appears in the
architectural description, yet the fusion score is the most discriminative single
quantity in the system, with an AUC of $0.925$ against $0.695$ for the semantic
score. Downstream of these, three enforcement states are emitted: \texttt{block}
for 10,604 prompts, \texttt{allow} for 14,364 and \texttt{allow\_limited} for 32.
Binary metrics count \texttt{block} as positive and both allow states as
negative, and since all 32 \texttt{allow\_limited} prompts are malicious, the
intermediate state contributes only false negatives here. The mapping is the
gateway's own: \texttt{allow\_limited} ends in the same final status as
\texttt{allow}, the orchestration layer executes the request, and we found no
restriction attached to it anywhere in the code. It is given to an unescalated
review, and to an allow close to the review threshold, when the residual evidence
is weak; Section~\ref{sec:results-mapping} reports the main runs with the state
counted as a block instead. Because human review is
disabled, any input left in a review state is converted to \texttt{block}
unless the enforcement rule releases it as \texttt{allow\_limited}, and
this fail-closed conversion is the single largest source of true positives
attributed to the arbitration tier (Section~\ref{sec:results-ablation}).

The supporting machinery diverges as well. Of eighteen components in the
middleware package, six run during evaluation---request tracing, obfuscation
decoding, provenance labeling, similarity memory, calibration bridging and
drift detection---while the remaining twelve are skipped, being disabled in
evaluation mode, acting on tool responses rather than prompts, or not imported by
the pipeline at all. The trust boundary among them assigns a string label and
validates it against a fixed set, with no HMAC, signature or key material
involved; since none of the five corpus source names belongs to the recognized
set, all 25,000 prompts receive the same label, \texttt{UNTRUSTED\_EXTERNAL}, and
the mechanism has no discriminative variance in this evaluation. One further
component appears in no design document at all: a similarity memory that matches
each input against 118 stored patterns by Jaccard similarity over word tokens,
referring at one threshold and blocking outright at another, and accounting for
2,132 hard blocks.

A further enforcement step is absent from every design document. When the policy
stage returns \emph{allow}, the enforcement stage may still escalate the request
to a block on the strength of risk signals carried forward from earlier tiers;
we call this the post-allow escalation. It is not a tier, has no name in the
architecture, and was not isolated by the first series of runs; disabling it
(Section~\ref{sec:results-postallow}) costs more detections than deleting the
arbitrator.

\paragraph{Terms} Four names recur below. The \emph{rescue branches} are five
hand-written policy rules, each named after the failure mode it was added to
recover, that send inputs to review (Section~\ref{sec:results-decides}). The \emph{allow-hint rail} is the
gate that may turn a benign arbitrator verdict into an allow. The
\emph{post-allow escalation} is the enforcement step, described above, that
blocks some inputs the policy stage allows. \texttt{allow\_limited} is the
enforcement state given to an unescalated review, or to an allow close to the
review threshold, when the residual evidence is weak; like \texttt{allow}, it
ends in an executed request.

\subsection{Divergences, and why they matter here}\label{sec:system-divergence}

Table~\ref{tab:divergence} collects the differences. We list them not as
implementation defects---several are ordinary drift between a design document and
a codebase that outgrew it---but because each changes what a component ablation
performed on this system can be taken to mean.

\begin{table*}[t]
\caption{Documented design against instrumented behavior. Each row is
established from the traces of the control run, or from the source revision that
produced them.}
\label{tab:divergence}
\centering
\small
\begin{tabular}{p{0.44\textwidth}p{0.48\textwidth}}
\toprule
\textbf{As documented} & \textbf{As measured} \\
\midrule
Semantic layer scores against 32 verified attack-category centroids &
23 category references and 192 support examples (215 texts) \\
\addlinespace
Adaptive percentile calibrator with IQR outlier rejection and poisoning
guardrails &
Two review values keyed on a boolean risk flag; constant block threshold;
order-invariant under a full permutation; an undocumented per-user reputation
rule, inactive in the control \\
\addlinespace
Neural arbitrator resolves borderline cases &
Arbitrator receives a band already $96.82\%$ malicious; its verdicts decide at
most 152 of 25,000 decisions \\
\addlinespace
Deterministic allow-hint rail prevents the model from clearing a threat &
Applied zero times in the control; when the verdict is forced benign for the
whole band, the margin condition rejects 7,441 of 7,453 \\
\addlinespace
Allow-hint gated by four conditions &
Eight conditions besides an enable flag; the semantic-margin condition alone fails in 2,099 of 2,100
checks \\
\addlinespace
Enforcement maps the policy outcome &
An undocumented post-allow escalation converts 1,221 policy allows into blocks
(1,146 malicious) \\
\addlinespace
Cryptographically verified provenance tags &
String label validated against a fixed set; no HMAC or signature \\
\addlinespace
Untrusted tool output receives elevated scrutiny &
All 25,000 prompts receive the same \texttt{UNTRUSTED\_EXTERNAL} label \\
\addlinespace
Ten-component middleware stack &
Eighteen components present, six run during evaluation \\
\addlinespace
Four-tier cascade &
Four tiers plus an undocumented seven-signal fusion stage, a ten-reason policy
stage and a similarity memory \\
\bottomrule
\end{tabular}
\end{table*}

Three rows bear directly on the measurements that follow, and they correspond to
the three questions the rest of the paper answers. The first is that the
arbitration tier does not receive borderline traffic in the sense the design
intends, since routing delivers it a nearly pure malicious population, which is
the condition under which the base-rate argument of Section~\ref{sec:baserate}
applies. The second is that the allow-hint rail is inert, and a design in which a
neural verdict can escalate but almost never de-escalate makes the tier's output
nearly one-directional, with consequences measured in
Section~\ref{sec:results-repair}. The third is that the components absent from
the architectural description---the fusion stage, the policy branches, the
similarity memory---are the ones carrying the detections
(Section~\ref{sec:results-decides}); an ablation confined to the four named tiers
cannot locate them, which is the practical reason a component ablation and a
substitution intervention on this system return different answers.

None of this is visible in any aggregate metric the system reports. Its recorded
accuracy, precision, recall and F1 are all reproduced exactly by the control run
of Section~\ref{sec:results-control}, with the arbitration tier receiving
corrupted input throughout.

\section{Method}\label{sec:method}

This section sets out the four operations we apply to a pipeline component,
introduces the quantity we propose reporting beside them, and states how the runs
were carried out.

\subsection{Deletion, substitution, repair, transplant}\label{sec:three-interventions}

Write $P$ for a gateway mapping an input $x$ to an enforcement outcome, and $c$
for one of its components. An ablation study is normally read as answering how
much of $P$'s performance $c$ supplies. Four distinct operations all travel
under the name ablation, however, and they answer different questions.

Under \emph{deletion}, the component is switched off and its traffic falls
through to whatever handles the residue, giving $P_{\neg c}$. The quantity
$P - P_{\neg c}$ mixes the component together with everything that exists only
on its account---the predicate selecting inputs for it, the score interval
defining its remit, and the rule that fires when it yields no answer---and this
is the operation behind the ablation tables surveyed in
Section~\ref{sec:related}. Under \emph{substitution}, by contrast, the component
is called as usual, but whatever it returns is overwritten by a constant $k$,
giving $P_{c \leftarrow k}$; selection predicate, interval and residue rule all
stay in place, so $P - P_{c \leftarrow k}$ captures the influence of the
component's \emph{answers} and nothing else, and running both extreme constants
brackets that influence from either side. The interpretability literature calls
the analogous operation an interchange or patching intervention
\cite{geiger2021,vig2020}, and statisticians have long held that dropping a
covariate measures something other than its importance \cite{hooker2021,lei2018}.
Under \emph{repair}, a fault in the component's implementation is
corrected and nothing else is touched, giving $P_{c \rightarrow c'}$, so that the
difference exposes how strongly $P$'s output tracks $c$'s \emph{quality}; this
operation is available only where a component is genuinely broken and the break
genuinely mendable, and we could apply it because the arbitration stage turned
out to be running on corrupted tokenization (Section~\ref{sec:results-repair}).
Under \emph{transplant}, finally, the component is replaced by a different and
independently validated one, $P_{c \rightarrow d}$, with routing, thresholds and
enforcement again left in place. Repair can only be applied to a component that
happens to be broken; transplant can be applied to any stage that has a
replacement, and the difference $P_{c \rightarrow d} - P$ shows whether the
system would track a better component at all
(Section~\ref{sec:results-transplant}).

Reporting only the first invites the inference that a component is worth whatever
its absence costs. Section~\ref{sec:results} reports all four for one gateway,
over one corpus.

\subsection{Base rate of the arbitration band}\label{sec:baserate}

A cascade forwards only a slice of its traffic to its costliest stage. Call
$B \subseteq X$ the \emph{arbitration band}, the inputs that reach the
arbitrator, and let
\begin{equation}
p \;=\; \frac{\left|\{x \in B : y(x) = \text{malicious}\}\right|}{|B|}
\label{eq:baserate}
\end{equation}
be the malicious share inside it.

Two things follow at once. A stub that condemns everything without looking
achieves precision $p$ on $B$; and a real classifier can beat that stub only
through the inputs it \emph{spares}, since sparing a benign input turns a false
positive into a true negative while sparing a malicious one turns a true positive
into a false negative. The exchange pays only when the spared set is mostly
benign, so that as $p$ approaches unity there is scarcely any benign material
left to spare, and no classifier, at any level of accuracy, can improve on the
stub. The stage's ceiling is therefore fixed by its input distribution before its
own quality enters the account at all. Formally, against the constant-positive
stub a verdict can gain at most the number of benign inputs in the band,
$(1-p)|B|$, and only if each of its negative verdicts is both correct and
enforced as an allow; for the control run this is 237 of 25,000 decisions. The
bound limits what a verdict can gain over the stub; it does not limit what a
verdict can lose.

Axelsson's base-rate argument \cite{axelsson2000} runs the other way, observing
that when attacks are rare, false alarms swamp a detector's utility. Here routing
has manufactured the opposite extreme inside the band, and the consequence is
that the arbitrator's accuracy stops mattering; Sommer and Paxson's gap between
what a detector outputs and what an operator can do with it \cite{sommer2010} is
the same phenomenon seen from deployment.

Hence our proposal: \textbf{whenever a layered defense publishes an ablation
figure for a stage that receives routed traffic, it should publish the class
balance of that stage's band alongside it.} Lacking that number, a reader cannot
separate work done by the stage from work already finished upstream.

\subsection{Runs}\label{sec:protocol}

Three properties of \sysname{} (Section~\ref{sec:system}) make it usable as a
measurement object: it behaves deterministically once configuration and input
order are fixed, it can be made to emit its intermediate state at every stage,
and a recorded evaluation of it exists whose confusion matrix can be targeted for
exact reproduction. That last property is what turns the protocol below into a
falsifiable one, since a control run either lands on the recorded matrix or it
does not.

\paragraph{Corpus} Every run consumes the gateway's own benchmark of
$N = 25{,}000$ prompts, balanced 12,500 malicious against 12,500 benign, whose
SHA-256 sum is checked before each run begins. Of these, 20,519 come from the
project's in-house generator and 4,481 from four public collections
\cite{wildteaming,injecagent,llmailinject,yi2025}. %
How each public record relates to its original collection is established in
Section~\ref{sec:results-labels}; the short version is that only one of the four
enters the corpus with its texts unaltered, apart from the field names written
before each part.
The public records were not added to the generator's output but substituted for
part of it. The generator's code, run again, reproduces each of the 20,519
synthetic prompts exactly, and the 4,481 synthetic prompts that the public
records displaced from its earlier 25,000-prompt output were all of easy or medium
difficulty, so that $97.1\%$ of the synthetic attacks remaining are labeled hard
or expert. The step that selected them is not in the project's repository.

\paragraph{Configurations} Fifteen configurations, named in
Table~\ref{tab:runs}, were run end to end, each pushing all 25,000 prompts
through the live pipeline object. None was evaluated by replaying stored outcomes
offline, because the components we intervene on sit upstream of the policy stage,
and replaying recorded policy states would hold that stage constant by
construction, which is exactly the confound under examination.

\begin{table}[t]
\caption{The fifteen configurations. Each row is one end-to-end run over all
25,000 prompts. ``Fixed'' marks runs with the corrected tokenizer.}
\label{tab:runs}
\centering
\small
\setlength{\tabcolsep}{3pt}
\begin{tabular}{llc}
\toprule
\textbf{Label} & \textbf{Operation} & \textbf{Fixed} \\
\midrule
\textsc{Base}        & control, no switch set               & no  \\
\textsc{Sub-Pos}     & verdict pinned to ``malicious''       & no  \\
\textsc{Sub-Neg}     & verdict pinned to ``benign''          & no  \\
\textsc{Rem-Arb}     & arbitrator off, band fails open       & no  \\
\textsc{Rem-Mem}     & similarity memory off                 & no  \\
\textsc{Rem-Cal}     & calibration stage off                 & no  \\
\textsc{Rem-Resc}    & rescue policy branches off            & no  \\
\textsc{Core}        & \textsc{Sub-Pos} $+$ \textsc{Rem-Mem} $+$ \textsc{Rem-Cal} & no \\
\textsc{Ord-1}       & input order permuted                  & no  \\
\textsc{Ord-2}       & order permuted, calibration off       & no  \\
\textsc{Cpu}         & CPU-only execution provider           & no  \\
\midrule
\textsc{Repair}      & control, no switch set                & yes \\
\textsc{Repair-Pos}  & verdict pinned to ``malicious''       & yes \\
\textsc{Repair-Neg}  & verdict pinned to ``benign''          & yes \\
\textsc{Repair-Resc} & rescue policy branches off            & yes \\
\bottomrule
\end{tabular}
\end{table}

\paragraph{Switches} Each operation is triggered by an environment variable added
to the implementation, and with no variable set the pipeline must
land on the recorded confusion matrix exactly; that requirement is the control
establishing that the instrumentation is inert. The eleven runs without the
tokenizer correction execute one code state (a source commit plus a recorded
diff adding the switches) and the four with it a second (a later commit plus a
recorded diff carrying the correction); both diffs belong to the source code,
which is not part of the released bundle (see Data and Code Availability), and
each state passes the control separately.

\paragraph{Per-request traces} Every run writes one trace line per prompt, 51
fields wide, recording the corpus index and source, every stage's score, verdict
and elapsed time, the thresholds active for that prompt, the arbitrator's answer,
the outcome of each de-escalation condition taken separately, and the enforcement
state finally reached. In the control and its corrected twin, the arbitrator is
additionally scored on every prompt as a side computation kept out of the
decision path and out of all timing.

\paragraph{Mutable state} Several files are read and written by the pipeline, so
each run restores them from a snapshot beforehand and records their SHA-256 sums
on either side. One run shows a changed sum; Section~\ref{sec:limitations}
explains why and shows that its decisions were not affected.

\paragraph{Checking} A checking script, written before the runs and never edited
afterwards and part of the withheld source, audits each delivered batch for trace count and field completeness,
for agreement between traces and the summary files derived from them, for whether
the control lands on the recorded matrix, and---for each operation---for whether
the run's timings or verdicts were copied from the control instead of measured.
It also computes the exact McNemar statistic against the control. Every batch
reported here clears it.

\paragraph{Timing} Latency is taken only from runs executed one at a time on an
otherwise idle machine, after warming up on fifty texts held outside the corpus.
Peak resident memory is sampled once a second.

\subsection{Second series, external baselines and provenance}\label{sec:method-v3}

\paragraph{Second series} Twelve further runs on the second code state, each
over all 25,000 prompts and executed one at a time: the control three times; the
arbitrator's verdict pinned to ``malicious''; the post-allow escalation switched
off by a single added environment variable; and seven operating points
combining a semantic threshold between $0.30$ and $0.42$ with the allow-hint rail
on or off. The first run is gated: it must reproduce the first
series' control sample for sample, or no further run starts. It did.

\paragraph{Lineage groups} Many prompts in the corpus are near-copies of one
another, a text placed inside different wrappers or lightly reworded. Any
resampling or splitting that treats them as independent overstates precision and
leaks across splits \cite{kapoor2023}. We therefore group prompts by MinHash
locality-sensitive hashing over word 5-gram shingles \cite{broder1997} (128
permutations, 32 bands of 4), confirming each candidate pair by exact Jaccard
similarity $\geq 0.8$, and take connected components. This yields 21,807 groups,
the largest of 92 prompts, none mixing labels. All splits below are made at group
level, and all intervals in Sections~\ref{sec:results-external}
and~\ref{sec:results-labels} are percentile bootstrap intervals over resampled
groups ($B=1{,}000$).

\paragraph{External detectors} Four public detectors are scored on every record:
the Prompt Guard~2 classifiers at 86M and 22M parameters \cite{promptguard2card},
ProtectAI's DeBERTa-v3 prompt-injection classifier \cite{protectaiv2}, and Llama
Guard~3 at 1B parameters \cite{llamaguard3card}. Each is pinned to a recorded model
revision. The three classifiers accept 512 tokens; longer records are split into
consecutive windows and scored by the maximum over windows, as the Prompt
Guard~2 model card recommends. Llama Guard~3 is scored by the probability of
\texttt{unsafe} against \texttt{safe} at the first content token of its response
under its own chat template, the user text being truncated, never the
instruction; the two tokens carry at least $99.5\%$ of the probability mass on
every record, which confirms that the template is being followed.

\paragraph{Shortcut baselines} To see how much of the corpus can be solved from
surface form alone, four trivial models are fitted under group-level five-fold
cross-validation: the first word of the prompt, the logarithm of its length, a
character n-gram model over its first 64 characters, and a word n-gram TF-IDF
model over the whole text.

\paragraph{Operating points chosen away from the test data} Every detector,
\sysname{} included, has its operating point chosen by one rule: the highest
recall subject to a false positive rate of at most $5\%$ on the selection data.
For \sysname{} the candidates are the eight operating points recorded in the
second series (the control and the seven above); for the others they are score
thresholds. Two regimes are reported. In
the first, selection uses four fifths of the lineage groups and scoring the
remaining fifth, repeated over five fold assignments. In the second, selection
uses only the synthetic subset and scoring only the public records, which is the
situation of a defense calibrated on its own generator and deployed elsewhere.

\paragraph{Provenance} Every public record is traced back to its original
collection: InjecAgent and LLMail-Inject records by exact match after whitespace
normalization \cite{injecagent,llmailinject}, BIPIA records by locating each
of the benchmark's 250 published attack strings inside them \cite{yi2025}, and
WildJailbreak records by exact match against the released training split, which
recovers each record's original type \cite{wildteaming}.

\paragraph{Independent check set} A further 3,003 records are drawn directly
from the original collections, none of them in the corpus: the 203 benign
e-mails LLMail-Inject releases for false-positive testing; 400 original BIPIA
contexts (e-mail, table and code tasks) with the user's question, each once clean
and once with a BIPIA test attack inserted at its start, middle or end; and 500
WildJailbreak prompts of each of its four types. \sysname{} processes them twice,
once from a fresh state and once appended after the 25,000-prompt corpus so that
its state is that of a gateway in service.

\paragraph{Transplant runs} Four further runs on the second platform put an
external classifier into one or both of two stages, over the 25,000 corpus
prompts followed by the 3,003 check-set records, each against that platform's
own control. The classifier's score for each prompt is computed beforehand, as in
Section~\ref{sec:results-external}, and looked up when the stage is reached; its
threshold is the one selected on the synthetic subset alone. In the arbitration
stage the verdict is ``malicious'' when the score reaches that threshold and
``benign'' otherwise, and everything downstream of the verdict, including the
de-escalation conditions, is unchanged. In the semantic stage the classifier's
score replaces the combined semantic risk score through a piecewise-linear map
that sends its threshold to the control's semantic threshold of $0.42$, so that
routing and calibration act on it exactly as they acted on the original score.
The runs are Prompt Guard~2 (86M) in the arbitration stage, in the semantic
stage, and in both, and ProtectAI v2 in both. Environment variables select each
stage, and with none set the added code is never entered.
With none set, the instrumented revision reproduces the control run decision for
decision on all 28,003 prompts. Below the decision, 16 semantic scores differ by at
most $0.004$, and the one prompt of the residual non-determinism noted in
Section~\ref{sec:limitations} takes the other of its two memory scores and is
blocked by the arbitrator instead of by the memory.

\paragraph{Second platform} The control is re-executed on a second machine with a
different processor architecture, from the same source revision, model files and
state snapshots, all checked by SHA-256, and with the package versions of the
original environment.

\subsection{What the protocol does and does not settle}\label{sec:repro}

Everything in the two run series comes off one machine (an Apple-silicon
laptop running macOS on arm64, Python 3.13)
and two recorded code states, with corpus and model files checksummed at each
start; the independent re-execution of Section~\ref{sec:results-platform} uses
a second machine (x86-64, Windows) and the second code state. The
released bundle carries the traces for all fifteen runs, the
verification script that recomputes the main tables with its output, and the state snapshots (two of them by
digest only), so the tables in
Section~\ref{sec:results} can be rebuilt without touching the pipeline
\cite{barbican2026}.

Three things the protocol leaves open. Timing and memory describe this hardware
and generalize nowhere; no rival gateway was rebuilt under these conditions, and
the external classifiers of Section~\ref{sec:results-external} are baselines
rather than competitors, so nothing here ranks complete systems against one
another (the two defenses of Section~\ref{sec:results-others} are run with
library defaults, which no deployment need keep); and the benchmark belongs to the
gateway itself, which is a property whose consequences for generalization we
measure (Section~\ref{sec:results-corpus}) rather than attach afterwards as a
disclaimer.

\section{Results}\label{sec:results}

Unless stated otherwise, figures below are computed from the traces of the fifteen runs of
Table~\ref{tab:runs}. Intervals are Wilson $95\%$ intervals; paired comparisons
against the control use the exact McNemar test. Both treat the 25,000 prompts as
independent; near-copies (Section~\ref{sec:method-v3}) make the intervals and
$p$-values optimistic, though not the counts. Table~\ref{tab:results} collects
the outcome of every run in one place, and the subsections below take them in
turn: first the control that licenses the rest, then the disagreement between
deletion and substitution, then the repair that inverts the expected direction,
then the question of where the detections actually originate, and after it
the stage's cost, the corpus and the operating point, the second series, the
corpus labels, the external comparison, the check set, the transplants, two other
defenses, our earlier gateway and a second platform.

\subsection{Control}\label{sec:results-control}

With every switch unset, \sysname{} lands on its recorded confusion matrix:
$TP=10{,}084$, $FP=520$, $TN=11{,}980$, $FN=2{,}416$, giving recall $80.67\%$
$[79.97, 81.35]$, false positive rate $4.16\%$ $[3.82, 4.52]$, precision
$95.10\%$ and a Matthews correlation coefficient of $\mathrm{MCC}=0.774$. No prompt differs from that record.
Everything that follows is a departure from this run.

\begin{table*}[t]
\caption{All fifteen runs. $b$ and $c$ are discordant counts against the control
($b$: control correct, run wrong; $c$: the reverse); $p$ is the exact McNemar
probability.}
\label{tab:results}
\centering
\small
\begin{tabular}{lrrrrrrrrr}
\toprule
\textbf{Run} & \textbf{TP} & \textbf{FP} & \textbf{FN} & \textbf{Recall \%} &
\textbf{FPR \%} & \textbf{F1 \%} & \textbf{MCC} & $b/c$ & $p$ \\
\midrule
\textsc{Base}        & 10,084 & 520 & 2,416 & 80.67 & 4.16 & 87.29 & 0.774 & --- & --- \\
\textsc{Sub-Pos}     & 10,096 & 520 & 2,404 & 80.77 & 4.16 & 87.35 & 0.775 & 0/12 & $5\times10^{-4}$ \\
\textsc{Sub-Neg}     & 9,944 & 520 & 2,556 & 79.55 & 4.16 & 86.61 & 0.764 & 140/0 & $1\times10^{-42}$ \\
\textsc{Rem-Arb}     & 9,411 & 519 & 3,089 & 75.29 & 4.15 & 83.91 & 0.727 & 673/1 & $2\times10^{-200}$ \\
\textsc{Rem-Mem}     & 9,826 & 314 & 2,674 & 78.61 & 2.51 & 86.80 & 0.775 & 258/206 & 0.018 \\
\textsc{Rem-Cal}     & 9,269 & 461 & 3,231 & 74.15 & 3.69 & 83.39 & 0.723 & 815/59 & $6\times10^{-171}$ \\
\textsc{Rem-Resc}    & 9,147 & 502 & 3,353 & 73.18 & 4.02 & 82.60 & 0.710 & 937/18 & $4\times10^{-250}$ \\
\textsc{Core}        & 8,903 & 255 & 3,597 & 71.22 & 2.04 & 82.21 & 0.718 & 1,193/277 & $2\times10^{-135}$ \\
\textsc{Ord-1}       & 10,084 & 520 & 2,416 & 80.67 & 4.16 & 87.29 & 0.774 & 0/0 & 1 \\
\textsc{Ord-2}       & 9,269 & 461 & 3,231 & 74.15 & 3.69 & 83.39 & 0.723 & 815/59 & $6\times10^{-171}$ \\
\textsc{Cpu}         & 10,083 & 520 & 2,417 & 80.66 & 4.16 & 87.29 & 0.774 & 2/1 & 1 \\
\midrule
\textsc{Repair}      & 10,043 & 520 & 2,457 & 80.34 & 4.16 & 87.09 & 0.771 & 48/7 & $1\times10^{-8}$ \\
\textsc{Repair-Pos}  & 10,096 & 520 & 2,404 & 80.77 & 4.16 & 87.35 & 0.775 & 0/12 & $5\times10^{-4}$ \\
\textsc{Repair-Neg}  & 9,945 & 520 & 2,555 & 79.56 & 4.16 & 86.61 & 0.764 & 139/0 & $3\times10^{-42}$ \\
\textsc{Repair-Resc} & 9,147 & 502 & 3,353 & 73.18 & 4.02 & 82.60 & 0.710 & 937/18 & $4\times10^{-250}$ \\
\bottomrule
\end{tabular}
\end{table*}

Two identities in the table are worth noting immediately. \textsc{Repair-Pos}
matches \textsc{Sub-Pos} exactly, as it must, since pinning the verdict bypasses
the classifier and repairing the classifier therefore changes nothing;
\textsc{Repair-Resc} matches \textsc{Rem-Resc} exactly, which is not forced by
construction, and means that with the rescue branches disabled the arbitrator's
quality ceases to affect any of the 25,000 outcomes.

\subsection{Deletion and substitution disagree by a factor of 4.8}
\label{sec:results-ablation}

Deleting the arbitration stage (\textsc{Rem-Arb}), which this gateway's switch
implements as failing open so that the policy stage allows the band's inputs,
costs 673 detections, recall dropping from $80.67\%$ to
$75.29\%$. Read as a component ablation, this says the arbitrator supplies 673 of
the system's detections, and it is the figure a conventional ablation table for
this gateway would carry.

Substituting a constant for its output says otherwise. Pinning the verdict to
positive (\textsc{Sub-Pos}) changes 12 decisions, all improvements, and pinning
it to negative (\textsc{Sub-Neg}) changes 140. The two sets do not overlap, and the arbitrator's own decisions
therefore account for at most 152 of 25,000 outcomes, or $0.61\%$, and the
interval between the two constants brackets its causal contribution from either
side. The remaining four fifths of the deletion figure belongs to the change of
status that deletion imposes. Under deletion the band's 7,453 inputs leave review,
so the fail-closed rule that blocks unresolved reviews no longer applies; the
post-allow escalation blocks 6,767 of them again, and the release rule for allowed
inputs lets 686 through, all but one malicious. Under negative substitution they
stay in review: the fail-closed rule blocks 7,301, the stricter release rule for
reviews lets 140 through, and the allow-hint gate 12. Both counts include the 12
inputs the control already releases as \texttt{allow\_limited}, which is why they
exceed the decisions that change, 674 and 140. Of the two substitutions, \textsc{Sub-Pos} is the
sharper result, since replacing a neural classifier with a one-line stub that
answers ``malicious'' without reading its input raises recall from $80.67\%$ to
$80.77\%$ ($p = 5\times10^{-4}$).

\subsection{Repairing the classifier makes the system worse}
\label{sec:results-repair}

The two measurements above bound what the stage's answers are worth but say
nothing about whether better answers would be worth more. The repair intervention
settles that, and it settles it in the unexpected direction.

The arbitration stage was running with a tokenizer misconfiguration, a
compatibility flag meant for a different model family having converted
SentencePiece word boundaries into unknown-token identifiers, inflating a
29-token input to 51 tokens and interposing \texttt{[UNK]} between words. Over
twenty probe texts the corrected inference tokenizer reproduces the training
tokenizer's identifiers exactly, $20/20$.

The repair works, in that on the evaluation corpus the arbitrator's AUC rises
from $0.392$---below chance---to $0.855$, and it now rejects $88.8\%$ of benign
inputs correctly. The system nonetheless gets worse: recall falls from $80.67\%$
to $80.34\%$ ($b=48$, $c=7$, $p=1\times10^{-8}$), with the false positive rate
unchanged at $4.16\%$, and the constant-positive stub remains the best of the
three. Across the four public corpora the repair changes recall by $0.00$
percentage points, the entire difference falling in the synthetic subset, where
recall drops from $86.45\%$ to $86.00\%$.

Table~\ref{tab:band} supplies the mechanism behind this inversion. Routing
forwards 7,453 prompts to the arbitrator, of which $96.82\%$ $[96.40, 97.19]$ are
malicious, against $30.11\%$ outside the band, so that discrimination has largely
happened before the arbitrator is consulted.

\begin{table}[t]
\caption{The arbitration band, before and after repair.}
\label{tab:band}
\centering
\small
\begin{tabular}{lrr}
\toprule
& \textsc{Base} & \textsc{Repair} \\
\midrule
Band size $|B|$                       & 7,453  & 7,452 \\
Band base rate $p$                    & 96.82\% & 96.82\% \\
Base rate outside the band            & 30.11\% & 30.12\% \\
Arbitrator AUC (corpus)               & 0.392  & 0.855 \\
Positive verdicts                     & 6,111  & 4,705 \\
Precision of those positives          & 96.33\% & 98.85\% \\
Precision of a constant-positive stub & 96.82\% & 96.82\% \\
Negative verdicts                     & 1,342  & 2,747 \\
\quad of which actually malicious     & 99.0\%  & 93.3\% \\
\quad released as \texttt{allow\_limited} & 12     & 53 \\ %
De-escalations applied                & 0      & 0 \\
\bottomrule
\end{tabular}
\end{table}

The table can be read as a single chain. In a band with $p = 96.82\%$ the
constant-positive stub is right $96.82\%$ of the time, whereas the broken
classifier's positive verdicts are right $96.33\%$ of the time, which is worse
than the stub. The repaired classifier's positives are more precise, at
$98.85\%$, but far fewer, since it issues them for $63.1\%$ of the band rather
than $82.0\%$; and because a positive verdict ends in a block in this
pipeline, by escalation or by the fail-closed rule, while a negative verdict produces an allow only through the
\texttt{allow\_limited} branch of the enforcement rule, coverage is what converts
into detections, so that escalations fall from 4,611 to 3,573 and 48 detections
are lost, every one of them an input released through that branch. The same repaired classifier's negative verdicts are wrong
$93.3\%$ of the time, which is not a defect of the classifier but a property of
the band: where $96.82\%$ of the population is malicious, almost any negative
verdict is wrong, however well calibrated the model. Only 53 of those 2,747
negative verdicts reach an allow decision in any case, all of them through the
\texttt{allow\_limited} branch: the de-escalation gate itself never passes, being
applied zero times in both runs, so that the gate which renders the classifier's
verdict inert is the same gate that protects the system from acting on the
classifier's errors.

The architecture, in short, converts classifier output into blocks and, through
one narrow branch, into a few dozen allows. Accuracy manifests as a willingness to say ``benign'', and saying
``benign'' has no effect, so in these two runs system recall falls as arbitrator
accuracy rises. The trend does not continue: the transplanted classifier of
Section~\ref{sec:results-transplant}, more accurate on the band than either,
raises recall slightly.

\subsection{Sensitivity to the enforcement mapping}\label{sec:results-mapping}

Every figure above counts \texttt{allow\_limited} as the gateway does, as an
allow (Section~\ref{sec:system-instrumented}). A deployment could nevertheless
attach a restriction to that state, and Table~\ref{tab:mapping} reports the main
runs under both readings.

\begin{table}[t]
\caption{Decisions changed against the control under the gateway's own mapping
(\texttt{allow\_limited} counted as an allow) and with \texttt{allow\_limited}
counted as a block, and how many of the former are moves into or out of
\texttt{allow\_limited}. Transplant runs (Prompt Guard~2) on the second
platform, against its control.}
\label{tab:mapping}
\centering
\small
\setlength{\tabcolsep}{4pt}
\begin{tabular}{lrrr}
\toprule
\textbf{Run} & \makecell[r]{as allow\\(reported)} & \makecell[r]{as\\block} & \makecell[r]{via\\\texttt{allow\_limited}} \\
\midrule
\textsc{Rem-Arb}                       & 674   & 0     & 674 \\
\textsc{Sub-Pos}                       & 12    & 0     & 12  \\
\textsc{Sub-Neg}                       & 140   & 12    & 128 \\
\textsc{Repair}                        & 55    & 0     & 55  \\
Transplant, arbitration               & 16    & 0     & 16  \\
\midrule
\textsc{Rem-Mem}                       & 464   & 451   & 13  \\
\textsc{Rem-Cal}                       & 874   & 894   & 0   \\
\textsc{Rem-Resc}                      & 955   & 962   & 5   \\
Post-allow off                         & 1,221 & 1,241 & 0   \\
Transplant, semantic                  & 1,015 & 1,057 & 4   \\
\bottomrule
\end{tabular}
\end{table}

The two readings agree on every component except the arbitrator. For the
memory, the calibration rule, the rescue branches, the post-allow escalation and
a semantic-stage transplant, the counts differ by at most 42 decisions and almost
none pass through \texttt{allow\_limited}. For the arbitrator every decision that
deletion, substitution, repair or transplant changes is a move into or out of
that state, apart from the 12 that negative substitution releases through the
allow-hint gate, and with the state counted as a block these operations change no
other decision. With the original semantic scores the arbitrator's effect on what
the gateway enforces thus runs almost entirely through releases with weak residual
evidence. Deletion releases 686 band inputs that way, through the rule for
allowed inputs, and negative substitution 140, through the stricter rule for
reviews, which produces the 673-to-140 discrepancy of decisions of
Section~\ref{sec:results-ablation}; under the restrictive reading only the 12
allow-hint releases remain, and the discrepancy disappears together with nearly
every effect of the stage. That the stage's quality does not reach
the output holds under both readings; how far deletion overstates the stage is a
property of the enforcement convention, which is why the convention has to be
reported with the figure (Section~\ref{sec:missed-recommendations}).

The base rate is not the only reason the stage is inert. On the independent
check set of Section~\ref{sec:results-check} the band holds 323 of the 2,003
real benign inputs and attacks and is $43.65\%$ malicious among them ($47.81\%$
over all 366 band inputs of the 3,003 records), so the ceiling of
Section~\ref{sec:baserate} does not bind; yet the transplanted classifier, which
labels 114 of the band's benign inputs benign, changes no decision there,
because all 114 are still blocked. Where the base rate does not make the stage
irrelevant, the enforcement asymmetry does.

\subsection{Where the detections come from}\label{sec:results-decides}

If the arbitrator's answers are worth at most 152 decisions, the 10,084 true
positives must originate elsewhere. Table~\ref{tab:marginal} attributes them to
components using the deletion and substitution runs already reported, its
rightmost column giving the precision of the blocks each component adds relative
to the configuration without it.

\begin{table}[t]
\caption{Marginal contribution of each component, against \textsc{Base}.}
\label{tab:marginal}
\centering
\small
\begin{tabular}{lrrr}
\toprule
\textbf{Component} & \textbf{+TP} & \textbf{+FP} & \textbf{Precision} \\
\midrule
Rescue policy branches      & 937 & 18  & 98.1\% \\
Calibration threshold rule  & 815 & 59  & 93.2\% \\
Review status (fail-closed) & 673 & 1   & 99.9\% \\
Similarity memory           & 258 & 206 & 55.6\% \\
Arbitrator decisions        & \multicolumn{3}{c}{$\leq 152$ decisions} \\
\bottomrule
\end{tabular}
\end{table}

Two entries deserve comment. The similarity memory adds blocks at $55.6\%$
precision, close to chance on a balanced corpus; and the rescue branches, which
are policy rules written by hand and each named after the failure mode it was
added to recover, contribute more detections than any other component.

Those branches are bound to a single generator. Although only $72.2\%$ of
malicious prompts are synthetic, four of them fire on synthetic prompts $100.0\%$
of the time and never fire on any of the four public corpora, and the mechanism
is not attack semantics but formatting: the branches match the literal strings
\texttt{Turn 2 - User:} and \texttt{[Context: Tool:}, which the project's
generator emits from its multi-turn and tool-context helpers, and which occur
3,269 and 2,823 times in the synthetic subset and zero times in the public
corpora. The branches, in effect, recognize which generator wrote the text. We
checked whether they were fitted to the evaluation corpus and found that they were
not, since all five predate that corpus by one to three weeks and the
1,000-prompt benchmark their commit messages cite is drawn from a different
corpus with $0\%$ text overlap with the evaluation set; the coupling is to the
generator family, not to the test split.

The second-largest contributor is described in the documentation as an adaptive
percentile calibrator with outlier rejection, and it too turns out to be
something simpler. In the control run its review threshold takes exactly two
values---$0.33$ when a risk signal is present, $0.36$ otherwise---and its block
threshold is constant at $0.49$ across all 25,000 prompts, while a full
permutation of the processing order (\textsc{Ord-1}) changes the threshold for
zero prompts and the decision for zero prompts ($b=c=0$); the same holds with the
stage disabled, \textsc{Ord-2} matching \textsc{Rem-Cal} exactly. The stage is
nonetheless consequential, since disabling it costs 815 detections. What its
ablation figure measures is a threshold rule, not adaptation---which is an
instance of the general point, an ablation number attaching to the name of a
component rather than to the mechanism actually doing the work.

Ranking single scores by AUC over the corpus completes the picture, placing the
undocumented fusion score ($0.925$) well above the semantic layer ($0.695$), the
rule layer ($0.595$) and the arbitrator ($0.392$, or $0.855$ after repair).

\subsection{What the arbitration stage costs}\label{sec:results-cost}

The arbitrator's forward pass, as recorded in the traces, averages $55.2$~ms over
all requests, $58.6\%$ of total pipeline latency, and pinning its verdict
(\textsc{Sub-Pos}) lowers the mean request by $61.4$~ms. Requests reaching it average $272.1$~ms against $18.7$~ms for
those that do not, while the median request costs $20.4$~ms and the 95th
percentile $281.0$~ms. Over the corpus the forward pass takes 1,381 seconds, in exchange for at most 152 changed decisions, and for a single
one on the four public corpora. %

\subsection{Corpus composition, leakage, and the operating point}
\label{sec:results-corpus}

The figures above are aggregates over a corpus that is not homogeneous, and the
aggregate conceals how unevenly the gateway performs across its parts.
Table~\ref{tab:sources} reports the control run by corpus source: recall ranges
from $99.59\%$ on BIPIA to $20.10\%$ on WildJailbreak, and three public sources
contain no benign prompts, so no false positive rate is defined for them.

\begin{table}[t]
\caption{\textsc{Base} by source. Repair changes none of these figures except the
synthetic row, where recall falls to $86.00\%$.}
\label{tab:sources}
\centering
\small
\begin{tabular}{lrrrr}
\toprule
\textbf{Source} & \textbf{mal.} & \textbf{ben.} & \textbf{Recall \%} & \textbf{FPR \%} \\
\midrule
synthetic       & 9,019 & 11,500 & 86.45 & 3.23 \\
BIPIA           & 981   & 0      & 99.59 & n/a \\
LLMail-Inject   & 1,000 & 0      & 92.70 & n/a \\
InjecAgent      & 500   & 0      & 36.40 & n/a \\
WildJailbreak   & 1,000 & 1,000  & 20.10 & 14.90 \\
\midrule
non-synthetic   & 3,481 & 1,000  & 65.70 & 14.90 \\
\bottomrule
\end{tabular}
\end{table}

Leakage accounts for little of the gap. The similarity memory holds 118 entries,
of which 96 appear verbatim in the evaluation corpus, matching 95 distinct
prompts, and removing them lowers recall from $80.67\%$ to $80.56\%$, so that
demonstrable leakage is worth $0.11$ percentage points. Removing instead every
prompt the memory matched at runtime ($n=4{,}126$) lowers recall to $73.84\%$,
but that subset measures the memory's whole contribution rather than leakage.
Neither is the dominant term: the larger effect is compositional, since on the
non-synthetic $17.9\%$ of the corpus recall is $65.70\%$ and the false positive
rate $14.90\%$.

An earlier episode in the project's audit history bears on the same question. The
history records 413 entries seeded into the similarity memory directly from the
evaluation corpus in August 2026, and an intermediate benchmark reporting
$86.89\%$ recall while they were present; they were removed five days later,
before the canonical run. We confirm that the canonical result is unaffected,
since re-executing with the cleaned 118-entry memory reproduces the recorded
matrix exactly, which it could not do had the 413 entries been active, and the
two figures are in any case not comparable, the contaminated run having used a
different routing threshold. What the episode illustrates is that the
contamination surfaced through a targeted audit and through nothing in the
aggregate metrics.

\label{sec:results-opt}
One further property of the evaluation belongs in the same accounting. The
configuration reported here was chosen over a predecessor by comparing both on
the same 25,000 prompts, with the stated aim of reducing that predecessor's false
positive rate, and no held-out split exists; in the taxonomy of Arp et al.\ this
is selective snooping \cite{arp2022}.
We have since re-run the selection on held-out partitions
(Section~\ref{sec:results-external}), and the inflation it introduces turns out
to be negligible: across eight recorded operating points the in-sample recall
ranges only from $80.67\%$ to $81.27\%$, and selecting among them on four fifths
of the lineage groups and scoring the fifth gives $81.3\%$ $[80.5, 82.0]$. The
snooping is real but cheap here; the operating point is not where the reported
figure's weakness lies.

\subsection{The second series: repetition, post-allow escalation, threshold}
\label{sec:results-postallow}

Table~\ref{tab:series2} reports the twelve runs of the second series. The
control reproduces the first series' control sample for sample, and its two
repetitions reproduce it again, down to every stored score: on one machine the
pipeline is deterministic.

\begin{table*}[t]
\caption{Second series, all runs over 25,000 prompts on the second code state.
$b/c$ and $p$ as in Table~\ref{tab:results}, against the control; band figures
are the size and malicious share of the arbitration band; latency is the mean and
95th percentile per request.}
\label{tab:series2}
\centering
\small
\begin{tabular}{lrrrrrrrr}
\toprule
\textbf{Run} & \textbf{TP} & \textbf{FP} & \textbf{Recall \%} & \textbf{FPR \%} &
$b/c$ & $p$ & \textbf{Band ($p_B$)} & \textbf{Latency ms} \\
\midrule
Control                       & 10,084 & 520 & 80.67 & 4.16 & --- & --- & 7,453 (96.82\%) & 89.8 / 231 \\
Control, repeat 1             & 10,084 & 520 & 80.67 & 4.16 & 0/0 & 1 & 7,453 (96.82\%) & 89.4 / 228 \\
Control, repeat 2             & 10,084 & 520 & 80.67 & 4.16 & 0/0 & 1 & 7,453 (96.82\%) & 89.6 / 229 \\
Verdict pinned ``malicious''  & 10,096 & 520 & 80.77 & 4.16 & 0/12 & $5\times10^{-4}$ & 7,453 (96.82\%) & 34.5 / 57 \\
Post-allow escalation off     & 8,938 & 445 & 71.50 & 3.56 & 1,146/75 & $8\times10^{-247}$ & 7,453 (96.82\%) & 88.8 / 226 \\
\midrule
Threshold $0.42$, rail on / off & 10,084 & 520 & 80.67 & 4.16 & 0/0 & 1 & 7,453 (96.82\%) & 89.0 / 227 \\
Threshold $0.36$, rail on     & 10,084 & 520 & 80.67 & 4.16 & 0/0 & 1 & 7,453 (96.82\%) & 89.4 / 228 \\
Threshold $0.34$, rail on / off & 10,084 & 520 & 80.67 & 4.16 & 0/0 & 1 & 7,453 (96.82\%) & 89.0 / 227 \\
Threshold $0.32$, rail on     & 10,100 & 520 & 80.80 & 4.16 & 0/16 & $3\times10^{-5}$ & 7,484 (96.83\%) & 89.1 / 227 \\
Threshold $0.30$, rail on     & 10,159 & 553 & 81.27 & 4.42 & 33/75 & $7\times10^{-5}$ & 8,019 (96.35\%) & 93.0 / 228 \\
\bottomrule
\end{tabular}
\end{table*}

Pinning the verdict to ``malicious'' repeats the first series' \textsc{Sub-Pos}
result exactly, 12 decisions changed and all of them corrections, and it removes
the arbitrator's forward pass: mean latency falls from $89.8$ to $34.5$~ms and the
95th percentile from $231$ to $57$~ms. The arbitrated requests themselves average
$261$~ms. Some three fifths of the gateway's mean latency is thus spent on a
stage whose answers are worth twelve decisions.

The largest single effect in either series belongs to a step that no tier owns.
Switching off the post-allow escalation of Section~\ref{sec:system-instrumented}
costs 1,146 detections and removes 75 false positives, recall falling from $80.67\%$ to
$71.50\%$: more than deleting the arbitrator (673), the calibration rule (815) or
the rescue branches (937). These are exactly the 1,146 and 75 inputs that the
first series could only describe. An ablation
confined to the named tiers would have attributed them to whichever tier had
produced the risk signals the escalation reads.

Here, unlike for the arbitrator, the deletion can be read at face value, and a
substitution shows why. The step sees the 15,605 requests that the policy stage
allows and the memory does not block outright, only $22.75\%$ of them malicious,
and the 1,221 it escalates are $93.86\%$ malicious. Computed from the control
traces, escalating a random 1,221 of the same requests would catch 278 attacks
(95\% interval 251--306) instead of 1,146 and raise the false positive rate to
$11.11\%$, and escalating all of them would reach $99.90\%$ recall at a false
positive rate of $100\%$. A constant is a poor substitute here because the step
receives a band in which the minority is the class it acts on. The two
operations agree for this step and disagree for the arbitrator, so the
disagreement of Section~\ref{sec:results-ablation} is a property of the
arbitrator's band, not of the method.

Executing the constant substitution on the second platform shows that these
figures hold only if nothing else changes. Escalating every allowed request,
including the 50 warm-up prompts, pushes the benchmark user past the reputation
rule of Section~\ref{sec:system-instrumented} before the first measured prompt,
and both calibration thresholds stay $0.10$ lower for all 25,000. The arbitration
band doubles to 14,944 prompts, $71.09\%$ malicious, and the run ends at
$99.40\%$ recall and a false positive rate of $96.90\%$; 13,934 decisions differ
from the control. A substitution executed in the running system thus measures
the step together with everything its output feeds; the figures computed from
the traces measure the step alone.

The threshold sweep confirms the picture of Section~\ref{sec:results-decides}
from the other side. Between $0.34$ and $0.42$ the semantic threshold changes no
decision at all, because the calibrated review threshold below it is what
binds; the allow-hint rail can be switched off without effect, because in all
2,100 checks it never passes, its semantic-margin condition alone failing in
2,099. Lowering the threshold to $0.30$ widens the arbitration band to 8,019
prompts, still $96.35\%$ malicious, and gains 75 detections for 33 new false
alarms.

\subsection{What the corpus labels encode}\label{sec:results-labels}

A per-source table (Table~\ref{tab:sources}) presumes that each source's records
mean what the source meant by them. Tracing the 4,481 public records to their
originals shows that this holds for one source in four.

\paragraph{InjecAgent} 490 of the 500 records match an original user instruction
and attacker instruction pair verbatim, and the remaining ten differ only in
format. This source enters the corpus as published, apart from the field names
written before each part (see below). %

\paragraph{LLMail-Inject} With whitespace and case normalized, the body of 948
of the 1,000 records occurs in the released challenge submissions, 884 of them
together with their subject; the remaining 52 we could not match. All 1,000
records carry appended lines absent from the original e-mails, among them the
challenge's outcome record for that submission, a JSON object with keys such as
\texttt{email.retrieved} and \texttt{exfil.sent}. Every malicious LLMail record
therefore ends in a string naming exfiltration. Removing the outcome line lowers
the two strongest external classifiers' recall on this source at the operating
points of Table~\ref{tab:external}, from $74.7\%$ to $66.2\%$ (ProtectAI) and
from $87.0\%$ to $81.8\%$ (Prompt Guard~2, 86M). Re-executing \sysname{} on the corpus with the line removed, on the second
platform of Section~\ref{sec:results-platform} and against that platform's own
control, changes its LLMail recall from $92.70\%$ to $92.50\%$, two decisions of 1,000,
and no decision on any other source ($b=2$, $c=0$, $p=0.5$). The line alters which
memory entry 108 of these prompts match best, but not the outcome: the gateway's
strength on this source is not borrowed from the appended metadata.

\paragraph{BIPIA} Each of the 981 records contains one of the benchmark's
published attack strings, but none contains a BIPIA context. The attacks are
embedded in generated HTML documents of median length 51,694 characters, a
format the original benchmark, whose contexts are e-mails, tables, code
discussions, news articles and abstracts, does not use. The source is better
described as BIPIA attacks inside synthetic web pages, and \sysname{}'s
$99.59\%$ on it says nothing about BIPIA itself; Section~\ref{sec:results-check}
measures that directly.

\paragraph{WildJailbreak} All 2,000 records are found in the released training
split, which assigns each prompt one of four types: plain harmful, plain benign,
adversarial harmful and adversarial benign, the last being prompts that use
jailbreak framing for a harmless request \cite{wildteaming}. The corpus label
follows form rather than harm. Its 1,000 ``benign'' records are the plain ones,
505 harmless and 495 harmful; its 1,000 ``malicious'' records are the adversarial
ones, 532 harmful and 468 harmless. The false positive rate of $14.90\%$ reported
for this source in Table~\ref{tab:sources} is therefore partly the blocking of
harmful requests, and the recall of $20.10\%$ partly the passing of harmless ones.

\paragraph{How the labels were assigned} The script in the project's repository
that collects the public records explains these findings. It labels every
WildJailbreak \texttt{vanilla} field benign and every \texttt{adversarial} field
malicious, reading no harm attribute. It takes the BIPIA-derived records not from
the benchmark's release but from a third-party repackaging on Hugging Face whose
dataset card is empty and which states no license \cite{bipiahf}, and it writes
each record as the page under the header \texttt{Context (HTML/Document):}
followed by the attack under the header \texttt{Potential injected
instruction:}. It writes InjecAgent records as field-name and value pairs, so
that 490 of the 500 contain the string \texttt{Attacker Instruction:}, and it
appends LLMail-Inject's outcome record under the header \texttt{Objectives:}.
All 981 BIPIA-derived records therefore announce their own label, as do 490
InjecAgent and 1,000 LLMail-Inject records, and none of these strings occurs in
any synthetic prompt.
Removing the four headers \texttt{Context (HTML/Document):}, \texttt{Potential
injected instruction:}, \texttt{User Instruction:} and \texttt{Attacker
Instruction:} from the BIPIA-derived and InjecAgent records and
re-executing \sysname{} on the second platform, against that platform's own
control, changes two decisions of 25,000, both InjecAgent attacks now passed
(recall $36.40\%$ to $36.00\%$; $b=2$, $c=0$, $p=0.5$). On the BIPIA-derived
records the first tier fires on 397 prompts instead of 462, but no final decision
changes and recall stays at $99.59\%$. Together with the LLMail-Inject result
above, the gateway's per-source figures do not rest on the strings that announce
the labels.

Table~\ref{tab:wjtypes} separates the four types. The external classifiers
respond to form: they flag adversarial prompts at $70$--$99\%$ whatever their
intent, and plain prompts rarely. Llama Guard~3 responds to harm: it flags $81\%$
of plain harmful requests and $12\%$ of harmless adversarial ones. \sysname{}
responds to neither. It flags harmless plain requests ($18.6\%$) more often than
harmful plain requests ($11.1\%$), and on the question it is built to answer,
separating adversarial harmful prompts from plain harmless ones, its fused score
reaches an AUC of $0.470$, below chance, where the ProtectAI classifier reaches
$0.991$.

\begin{table*}[t]
\caption{WildJailbreak records of the corpus by original type ($n$ per type in
parentheses). Flag rates use operating points selected on the synthetic subset
only (Section~\ref{sec:method-v3}). AUCs use each detector's continuous score
(the fused score for \sysname{}): \emph{attack} separates adversarial harmful from
plain benign, \emph{form} adversarial from plain, \emph{harm} harmful from
harmless.}
\label{tab:wjtypes}
\centering
\small
\begin{tabular}{lrrrrrrr}
\toprule
& \multicolumn{4}{c}{\textbf{Flagged \%}} & \multicolumn{3}{c}{\textbf{AUC}} \\
\cmidrule(lr){2-5}\cmidrule(lr){6-8}
\textbf{Detector} & \makecell[r]{plain\\benign (505)} & \makecell[r]{adv.\\benign (468)} &
\makecell[r]{plain\\harmful (495)} & \makecell[r]{adv.\\harmful (532)} &
attack & form & harm \\
\midrule
\sysname{}            & 18.6 & 16.2 & 11.1 & 24.1 & 0.470 & 0.443 & 0.524 \\
ProtectAI v2          &  0.6 & 70.5 &  3.2 & 71.2 & 0.991 & 0.981 & 0.556 \\
Prompt Guard 2, 86M   &  7.5 & 87.4 & 30.3 & 96.6 & 0.981 & 0.919 & 0.642 \\
Prompt Guard 2, 22M   &  6.9 & 96.6 & 30.7 & 99.4 & 0.987 & 0.926 & 0.622 \\
Llama Guard 3, 1B     &  5.3 & 11.8 & 81.2 & 52.8 & 0.924 & 0.522 & 0.897 \\
\bottomrule
\end{tabular}
\end{table*}

\paragraph{Surface form} The synthetic subset and the corpus as a whole can be
separated almost without reading. Under group-level cross-validation, the first
word of a prompt alone classifies $96.7\%$ of the corpus correctly (AUC $0.995$),
the first 64 characters $99.2\%$, and a bag of word n-grams $99.75\%$; every
public malicious record is identified by its first word, which names its source
format, and WildJailbreak's two halves are separated by length alone at an AUC of
$0.999$. A corpus in which the label is recoverable from a prompt's opening
word measures, at best, whether a detector has learned that word
\cite{geirhos2020}.

\subsection{Against public detectors}\label{sec:results-external}

Table~\ref{tab:external} sets \sysname{} beside the four external detectors,
every operating point chosen by the same rule on data the reported figure does
not use.

\begin{table*}[t]
\caption{\sysname{} against external detectors. AUC over all 25,000 records with
a group-bootstrap interval. \emph{In distribution}: operating point selected on
four fifths of the lineage groups, scored on the fifth. \emph{Under shift}:
selected on the synthetic subset, scored on the 4,481 public records (false
positive rate on the 1,000 WildJailbreak records labeled benign). Recall /
FPR in \%, intervals over resampled groups.}
\label{tab:external}
\centering
\small
\begin{tabular}{lcccc}
\toprule
\textbf{Detector} & \textbf{AUC} & \textbf{In distribution} & \textbf{Under shift} &
\textbf{LLMail recall} \\
\midrule
\sysname{} (recorded points) & ---                  & 81.3 [80.5, 82.0] / 4.4 & 65.8 [63.8, 67.9] / 14.9 & 92.7 \\
\sysname{} (fused score)     & 0.925 [0.921, 0.928] & 83.6 [82.8, 84.3] / 5.0 & 69.8 [67.9, 71.7] / 35.3 & 92.5 \\
ProtectAI v2                 & 0.962 [0.959, 0.964] & 75.1 [74.4, 75.9] / 5.0 & 80.5 [79.1, 82.0] / 1.9 & 74.7 \\
Prompt Guard 2, 86M          & 0.944 [0.940, 0.947] & 76.3 [75.5, 77.2] / 5.0 & 88.6 [87.4, 89.7] / 18.8 & 87.0 \\
Prompt Guard 2, 22M          & 0.827 [0.817, 0.837] & 48.3 [47.2, 49.4] / 5.1 & 74.5 [72.7, 76.1] / 18.7 & 64.2 \\
Llama Guard 3, 1B            & 0.904 [0.894, 0.912] & 50.6 [49.5, 51.7] / 4.8 & 50.0 [47.9, 52.0] / 42.9 & 34.1 \\
\bottomrule
\end{tabular}
\end{table*}

In distribution \sysname{} leads, by five points of recall over the best external
classifier at a slightly lower false positive rate. That is the result its own
evaluation would lead one to expect, and it is also the regime in which its
policy branches recognize the generator. Under shift the order reverses. The
ProtectAI classifier, calibrated on the same synthetic records, detects $80.5\%$
of the public attacks against \sysname{}'s $65.8\%$ while raising one eighth as
many false alarms ($1.9\%$ against $14.9\%$); it dominates on both axes. Prompt
Guard~2 at 86M detects more still, at a somewhat higher false alarm rate than
\sysname{}'s.
The one public source on which \sysname{} keeps its lead over all four is
LLMail-Inject; on the BIPIA-derived records every detector except Llama Guard~3
exceeds $98\%$.
Llama Guard~3, trained for harmful content rather than injection, is the weakest
injection detector here and the strongest harm detector of
Table~\ref{tab:wjtypes}, which is what its design predicts.

Two qualifications apply. The ProtectAI model card names only some of its
training collections, so overlap with WildJailbreak cannot be excluded; and the
false positive rate under shift rests on 1,000 benign records whose label means
``plain'' (Section~\ref{sec:results-labels}). The independent set of the next
subsection removes the second qualification but not the first, since its
WildJailbreak prompts come from the same released split.

\subsection{An independent check set}\label{sec:results-check}

Table~\ref{tab:check} reports the 3,003 records of the independent set. The
external detectors keep the operating points selected on the synthetic subset
(Table~\ref{tab:external}, \emph{under shift}); \sysname{} keeps its own. Its two
passes, from a fresh state and after the 25,000-prompt corpus, agree on every
decision, every fused score and every attack-memory similarity, so the corpus
leaves no trace in the state these prompts meet and a single column serves for
both. The first 25,000 decisions of the appended pass also reproduce the control
on the same machine exactly.

\begin{table*}[t]
\caption{Independent check set (3,003 records drawn from the original
collections, none in the corpus). Flag rates in \%. \emph{Real benign}: 203
LLMail-Inject false-positive e-mails, 400 clean BIPIA contexts and 500 plain
benign WildJailbreak prompts (1,103). \emph{Attacks}: the same 400 BIPIA
contexts with a test attack inserted, and 500 adversarial harmful WildJailbreak
prompts (900). AUC from each detector's continuous score.}
\label{tab:check}
\centering
\small
\begin{tabular}{lccccccc}
\toprule
 & \multicolumn{3}{c}{\textbf{Flag rate}} & & & \multicolumn{2}{c}{\textbf{AUC}} \\
\cmidrule(lr){2-4}\cmidrule(lr){7-8}
\textbf{Detector} & \makecell{LLMail\\benign} & \makecell{BIPIA\\clean} &
\makecell{BIPIA\\attacked} & \makecell{FPR, real\\benign} & \makecell{Attack\\recall} &
\makecell{BIPIA attacked\\vs clean} & \makecell{Attacks vs\\real benign} \\
\midrule
\sysname{}          & 1.5  & 28.8 & 34.3 & 18.0 & 31.2 & 0.510 & 0.547 \\
ProtectAI v2        & 0.0  & 17.3 & 15.8 & 6.5  & 46.3 & 0.502 & 0.872 \\
Prompt Guard 2, 86M & 0.0  & 36.0 & 56.3 & 16.9 & 78.2 & 0.627 & 0.906 \\
Prompt Guard 2, 22M & 0.0  & 6.3  & 8.3  & 6.4  & 58.3 & 0.639 & 0.831 \\
Llama Guard 3, 1B   & 40.9 & 38.3 & 47.0 & 25.6 & 49.6 & 0.559 & 0.716 \\
\bottomrule
\end{tabular}
\end{table*}

\paragraph{BIPIA} On original BIPIA contexts no detector separates the attacked
copy of a context from its clean copy: the paired AUCs lie between $0.50$ and
$0.64$, and \sysname{}'s is $0.510$. \sysname{} flags $34\%$ of attacked contexts
and $29\%$ of clean ones, and its clean flags fall mostly on code: it blocks
$61\%$ of the clean code-question contexts, against $12\%$ of clean e-mails and
$21\%$ of clean tables. The $99.59\%$ the corpus reports for BIPIA
(Section~\ref{sec:results-labels}) is therefore a property of the synthetic
pages that carry the attack strings, not of the attacks.

\paragraph{LLMail-Inject} On the 203 benign e-mails the competition releases for
false-positive testing, \sysname{} flags three ($1.5\%$) and the three
classifiers none. Llama Guard~3 flags $41\%$ of them, consistent with its being
a harm classifier rather than an injection detector.

\paragraph{WildJailbreak} The 2,000 fresh WildJailbreak prompts repeat
Table~\ref{tab:wjtypes}: \sysname{} flags $16.0$, $17.2$, $10.4$ and $28.8\%$ of
the plain benign, adversarial benign, plain harmful and adversarial harmful
prompts, and separates adversarial harmful from plain benign at AUC $0.497$,
against $0.978$--$0.992$ for the three classifiers.

\paragraph{InjecAgent in its own format} Outside Table~\ref{tab:check}, we also
ran the 1,054 published InjecAgent test cases in the form the benchmark uses,
the attacker instruction inside a simulated tool response, together with the 17
tool-response templates with the attack slot left empty. The cases carry the
same 62 attacker instructions as the corpus's InjecAgent records, there written
under field names; no text is shared with the corpus. \sysname{} detects
$45.4\%$ of these attacks, against $36.4\%$ of the corpus's InjecAgent records,
and flags 6 of the 17 clean templates. At the operating points of
Table~\ref{tab:check}, ProtectAI~v2 flags $95.4\%$ of the attacks and all 17
templates, Prompt Guard~2 86M $64.7\%$ and 3, Prompt Guard~2 22M $7.9\%$ and
none, and Llama Guard~3 $96.5\%$ and 8. LLM Guard flags $93.2\%$ and all 17,
its code scanner alone flagging every template. Seventeen templates are too few
for a false positive rate; they show only that ProtectAI~v2's and LLM Guard's
figures on this source are the flagging of tool responses as such.

Taken over all real benign and all attack records, \sysname{} has a false
positive rate of $18.0\%$ for a recall of $31.2\%$, and an AUC of $0.547$. Each of
the three injection classifiers is better on both axes, ProtectAI v2 for
instance at $6.5\%$ / $46.3\%$ and Prompt Guard~2 86M at $16.9\%$ / $78.2\%$;
Llama Guard~3 has the higher false positive rate but the higher recall and AUC.
None of these numbers is high in absolute terms; the BIPIA columns show that indirect
injection inside realistic documents is not solved by any detector tested here.

\subsection{Transplanting a stronger classifier}\label{sec:results-transplant}

The repair of Section~\ref{sec:results-repair} showed that the gateway does not
track the quality of its own arbitrator. A repaired classifier at $0.855$ ($0.789$
on the band) is
still a modest one, however, and it could be objected that a component good
enough would show through. Table~\ref{tab:transplant} answers the objection with
the external classifiers of Section~\ref{sec:results-external}, transplanted into
the arbitration stage, the semantic stage or both
(Section~\ref{sec:method-v3}).

\begin{table*}[t]
\caption{Transplant runs on the second platform, against that platform's control.
Corpus columns over the 25,000 prompts; $b/c$ and $p$ as in
Table~\ref{tab:results}. \emph{Band}: size and malicious share of the arbitration
band; \emph{band AUC}: the arbitrating classifier's score on that band.
\emph{Check set}: false positive rate on the 1,103 real benign records and recall
on the 900 attacks of Table~\ref{tab:check}; the last two rows repeat the
standalone classifiers from that table, at the same thresholds.}
\label{tab:transplant}
\centering
\small
\setlength{\tabcolsep}{4pt}
\begin{tabular}{lrrrrrrrr}
\toprule
& \multicolumn{6}{c}{\textbf{Corpus (25,000)}} & \multicolumn{2}{c}{\textbf{Check set}} \\
\cmidrule(lr){2-7}\cmidrule(lr){8-9}
\textbf{Run} & \makecell[r]{changed\\decisions} & $b/c$ & $p$ & \makecell[r]{Recall /\\FPR \%} &
\makecell[r]{Band\\($p_B$)} & \makecell[r]{band\\AUC} & \makecell[r]{FPR\\\%} & \makecell[r]{Recall\\\%} \\
\midrule
Control                          & ---   & ---         & ---                & 80.67 / 4.16 & 7,452 (96.82\%) & 0.368 & 18.0 & 31.2 \\
Prompt Guard 2 $\rightarrow$ arbitration & 16    & 4/12        & 0.077              & 80.74 / 4.16 & 7,453 (96.82\%) & 0.916 & 18.0 & 31.2 \\
Prompt Guard 2 $\rightarrow$ semantic    & 1,015 & 438/577     & $1\times10^{-5}$   & 81.31 / 3.69 & 8,182 (99.32\%) & 0.275 & 17.1 & 45.4 \\
Prompt Guard 2 $\rightarrow$ both        & 1,706 & 1,119/587   & $2\times10^{-38}$  & 75.94 / 3.68 & 8,182 (99.32\%) & 0.413 & 13.7 & 41.9 \\
ProtectAI v2 $\rightarrow$ both          & 2,988 & 1,369/1,619 & $5\times10^{-6}$   & 86.90 / 8.38 & 8,207 (88.80\%) & 0.714 & 15.2 & 50.9 \\
\midrule
Prompt Guard 2, standalone       & & & & & & & 16.9 & 78.2 \\
ProtectAI v2, standalone         & & & & & & &  6.5 & 46.3 \\
\bottomrule
\end{tabular}
\end{table*}

\paragraph{Arbitration stage} On the arbitration band Prompt Guard~2 reaches an
AUC of $0.916$, against $0.368$ for the gateway's own classifier on the same band.
Its positive verdicts are more precise ($99.38\%$ against $96.32\%$), and it
returns ``benign'' for 1,798 band inputs, 202 of them actually benign, where the
gateway's classifier returns it for 1,358 on this platform, of which 13 are
(1,342 on the first; Section~\ref{sec:results-platform}). Those 202 correct
verdicts are what a better arbitrator is for, and the pipeline releases none of
them: the false positive count stays at 520, the four inputs it releases are
attacks, and the run differs from the control in 16 decisions, 12 corrected and
four lost ($p = 0.077$). This is the
size of the constant-positive substitution of
Section~\ref{sec:results-ablation} (12 decisions), and on the check set not a
single decision changes. The base-rate argument of Section~\ref{sec:baserate}
therefore does not depend on the original classifier being broken; a competent
one is held to the same ceiling on what it can gain. It is not protected from
loss: with Prompt Guard~2 already in the semantic stage, placing it in the
arbitration stage as well changes 777 decisions, 724 of them lost, because the
changed semantic scores let the allow-hint gate pass 735 times.

\paragraph{Semantic stage} The same classifier in the semantic stage changes
1,015 decisions. Recall rises from $80.67\%$ to $81.31\%$ and the false positive
rate falls from $4.16\%$ to $3.69\%$; on WildJailbreak recall rises from $20.1\%$
to $45.0\%$ at a false positive rate of $15.7\%$ ($14.9\%$ before), and on the
check set recall rises from $31.2\%$ to $45.4\%$. The difference from the
arbitration stage is positional rather than qualitative. The semantic stage
decides what is routed, and its scores therefore reach the decision; the band it
leaves behind is $99.32\%$ malicious, more concentrated than before.

\paragraph{Both stages} With Prompt Guard~2 in both stages, recall on the corpus
falls to $75.94\%$, most of the loss on the synthetic subset ($86.5\%$ to
$77.8\%$), where the arbitrator's negative verdicts now reach an allow decision
735 times. ProtectAI v2 in both stages raises corpus recall to $86.90\%$ and
doubles the false positive rate to $8.38\%$, with WildJailbreak recall at
$73.6\%$ and InjecAgent at $76.8\%$ ($36.4\%$ in the control).

The check-set columns compare each transplant with the same classifier standing
alone at the same threshold. With Prompt Guard~2 in both stages, the gateway detects
$41.9\%$ of the attacks, against $78.2\%$ for the classifier on its own, at a
slightly lower false positive rate. ProtectAI v2 detects slightly more, $50.9\%$ against $46.3\%$, but
its false positive rate on real benign inputs rises from $6.5\%$ to $15.2\%$.
Either way the pipeline costs the classifier on at least one axis. In the
arbitration stage it reduces the classifier to the constant that the band's base
rate already implies; in the semantic stage it adds the false alarms of the
stages around it.

\subsection{Two other layered defenses}\label{sec:results-others}

Two open-source defenses were run on the check set, component by component, with
each component's own verdict recorded so that the stack and every deletion could
be computed afterwards. LLM Guard 0.3.16 \cite{llmguard} was run with its seven
input scanners for injection, invisible text, gibberish, language, code,
secrets and token limit, each at its library defaults; its injection scanner is
the ProtectAI v2 classifier of Table~\ref{tab:check} at a pinned revision
(89b085cd, against 90c9989b in our baseline) and at its default threshold of
$0.92$. NeMo Guardrails 0.24.1 \cite{nemo} was run with its three
jailbreak-detection checks, the length-per-perplexity and prefix/suffix
perplexity heuristics at their default thresholds ($89.79$ and $1845.65$) and
the embedding-plus-random-forest classifier; its LLM self-check rail was not
run. Both defenses flag an input when any component does, so neither routes:
every component sees every input. Table~\ref{tab:others} reports the result.

\begin{table}[t]
\caption{LLM Guard and NeMo Guardrails on the check set (1,103 real benign, 900
attacks; Table~\ref{tab:check}). \emph{Alone}: the component by itself;
\emph{deleted}: the stack without it. Rates in \%. Components that flag nothing
(invisible text, secrets, token limit; prefix/suffix perplexity) are omitted from
the rows and change nothing when deleted.}
\label{tab:others}
\centering
\small
\setlength{\tabcolsep}{4pt}
\begin{tabular}{lrrrr}
\toprule
& \multicolumn{2}{c}{\textbf{Alone}} & \multicolumn{2}{c}{\textbf{Deleted}} \\
\cmidrule(lr){2-3}\cmidrule(lr){4-5}
\textbf{Component} & FPR & Recall & FPR & Recall \\
\midrule
\multicolumn{5}{l}{\emph{LLM Guard, stack: FPR 24.1, recall 57.9}} \\
Injection (ProtectAI v2) & 1.5  & 32.6 & 23.8 & 26.6 \\
Code                     & 19.3 & 22.8 & 11.2 & 42.1 \\
Language                 & 3.4  & 2.7  & 21.0 & 55.8 \\
Gibberish                & 6.3  & 7.1  & 22.7 & 56.3 \\
\midrule
\multicolumn{5}{l}{\emph{NeMo Guardrails, stack: FPR 32.2, recall 94.9}} \\
Random forest            & 11.6 & 70.3 & 24.1 & 38.7 \\
Length/perplexity        & 24.1 & 38.7 & 11.6 & 70.3 \\
\bottomrule
\end{tabular}
\end{table}

\paragraph{Deletion without routing} In a stack where every component sees every
input, deleting a component removes exactly the inputs that it alone flags, so
deletion here measures what it should. What it measures is the difficulty. In
LLM Guard the code scanner is the only component flagging 142 attacks and also
the only one flagging 143 real benign inputs; in NeMo the perplexity heuristic
alone flags 221 attacks and 227 benign inputs. Both contributions come from format
detectors. The code scanner flags every BIPIA code context, attacked or clean, and
$90\%$ of the clean BIPIA e-mails, whose header line separates its fields with
vertical bars; the perplexity heuristic flags $99\%$ of the BIPIA tables, again
attacked or clean. A deletion table for either stack credits these components
with about a quarter of the recall, and the credit is for recognizing the
document type that the attacks of this set happen to be inserted into.

\paragraph{The classifiers} The two learned components separate the check set at
AUCs of $0.681$ (LLM Guard's injection scanner, at its revision) and $0.915$
(NeMo's random forest). Their per-category rates repeat the pattern of
Section~\ref{sec:results-labels}. NeMo's random forest flags $99.8\%$ of the
adversarial benign WildJailbreak prompts and $100\%$ of the adversarial harmful
ones, against $0.4\%$ of plain harmful and none of plain benign; LLM Guard's
injection scanner flags $50.4\%$ and $56.2\%$ of the two adversarial groups and
$2.6\%$ of plain harmful. Both respond to the adversarial form, which is also
what the corpus labels record. On BIPIA neither separates an attacked context
from its clean copy: NeMo flags $77\%$ of attacked and $73\%$ of clean e-mails,
LLM Guard's scanner $12\%$ and $15\%$.

\paragraph{On the corpus} Over the 25,000 corpus prompts LLM Guard's full stack
flags $46.3\%$ of the synthetic benign records, against $1.2\%$ for its injection
scanner alone; its recall is $99.7\%$ on BIPIA, $90.8\%$ on synthetic attacks,
$82.6\%$ on InjecAgent, $75.3\%$ on LLMail-Inject and $52.5\%$ on WildJailbreak.
The synthetic benign flags come from the language scanner ($36.6\%$; part of the
synthetic benign records are written in other languages) and the code scanner
($17.9\%$), not from the injection classifier.

NeMo Guardrails was run on a random sample of 3,000 corpus prompts, stratified
by source and label in proportion to the corpus (1,500 benign, 1,500 attacks),
because its perplexity checks take more than ten seconds per prompt on our
machine. On the sample it flags $1.1\%$ of the benign prompts and detects
$50.8\%$ of the attacks, against $32.2\%$ and $94.9\%$ on the check set. The
length-per-perplexity heuristic flags 116 of the 118 BIPIA attacks in the sample
and no benign prompt, and 180 of its detections are made by no other check; on
the check set the same heuristic flags $24.1\%$ of the benign inputs. The corpus
contains no benign document of the kind its BIPIA attacks are embedded in, so a
check that responds to the document looks like an attack detector there. On the
same 3,000 prompts LLM Guard flags $43.5\%$ of the benign prompts and detects
$86.0\%$ of the attacks.

\paragraph{Operating points} Both stacks were run at their library defaults. We
also chose thresholds for every component, keeping the highest recall at a false
positive rate of at most $5\%$. Chosen on the corpus, they do not carry over to
the check set: NeMo's reach $68.0\%$ recall at $4.9\%$ false positives on the
corpus sample and flag $41.6\%$ of the check set's real benign inputs; LLM
Guard's reach $60.0\%$ at $4.7\%$ on the corpus and $45.0\%$ at $8.4\%$ on the
check set. Chosen on the check set itself, in two-fold cross-validation
stratified by category over five seeds, which favors both defenses because a
clean BIPIA context and its attacked copy can fall into different folds, the
stacks reach $62.4\%$ (NeMo) and $39.7\%$ (LLM Guard) recall at false positive
rates of $4.9\%$ and $4.7\%$. Selecting the operating point on the corpus is
thus the larger risk: for NeMo it moves the false positive rate on real benign
inputs from $4.9\%$ to $41.6\%$.

Neither defense, then, offers a point of comparison that the base-rate argument
does not already cover. Their stacks are unrouted, so the ceiling of
Section~\ref{sec:baserate} does not apply and deletion reads unique
contributions correctly; but on data whose classes differ in form, the largest
unique contributions are made by components that read the form.

\subsection{A second routed system}\label{sec:results-cascade}

The ceiling of Section~\ref{sec:baserate} depends only on the class balance of
the input a routed stage receives, and the released records of our earlier
gateway \cite{cascade} allow that balance to be computed. In that system the
semantic stage sends inputs to review, and a local review model (Llama~3) is
consulted on part of the review band; its verdict can escalate a review to a
block, while its benign verdicts are held at review by a guard. Of the 5,000
records, 3,424 fall in the review band, $95.47\%$ of them malicious, and the
1,628 on which the model is consulted are $92.94\%$ malicious. The model's
malicious verdicts are $96.23\%$ precise, three points above a constant.

Counting only blocks as detections, the stage escalates 1,492 reviews to blocks.
A stub returning ``malicious'' for every consulted input differs from the model
in 136 decisions, 66 in the model's favor and 70 in the stub's (McNemar $p = 0.80$),
and the bound $(1-p)|B|$ on what any verdict could gain over the stub is 115
decisions. Counting reviews as detections, the stage changes no decision at
all. The gateway studied here is therefore not an isolated case: the second
cascade we could measure routes its costliest stage a band in which a constant
verdict is as accurate as the model, and what that stage contributes lies
outside accuracy, in the review queue it shortens.

\subsection{Re-execution on a second platform}\label{sec:results-platform}

Re-executing the control on an x86-64 machine from the same source revision,
model files and state snapshots reproduces the confusion matrix exactly,
$TP=10{,}084$, $FP=520$, $TN=11{,}980$, $FN=2{,}416$. Two of the 25,000 decisions
differ, in opposite directions, both on synthetic prompts whose arbitrator
probability lies within $0.06$ of the $0.5$ threshold. The rule layer is
identical on every prompt and the semantic scores differ by at most $0.0035$.
The arbitrator is not: its INT8 ONNX export returns a probability differing by
more than $0.01$ on 2,433 of the 7,452 prompts both platforms route to it, by
more than $0.1$ on 117, and in one case from $0.002$ to $0.992$ on identical
token sequences; 166 of its verdicts reverse. That 166 reversals yield two
changed decisions is the substitution result of
Section~\ref{sec:results-ablation} obtained a third way, by hardware rather than
by switch. Latency does not transfer: the arbitrator's own forward pass is
nearly unchanged, $179.6$~ms per arbitrated request against $176.0$~ms in the
second series' control, but the mean request takes $168.5$~ms
against $89.8$~ms, the semantic model running on a general-purpose processor
rather than on the original machine's accelerator, which every Mac run record
lists as the Core ML execution provider.

\section{Why the Aggregate Metrics Missed It}\label{sec:missed}

The measurements above turned up two faults in \sysname{}, one in a component and
one in how the evaluation was set up, and neither had left a trace in any number the
gateway reported. That is the more general finding, and it is worth separating
from the faults themselves: a reporting practice that cannot register a
completely inert neural stage, or a memory once seeded from the corpus it is scored on, is not
failing at the margin. This section examines how each fault escaped, and
Section~\ref{sec:missed-recommendations} states what would have caught them.
The corpus audit of Section~\ref{sec:results-labels} adds faults of the second
kind, label meanings altered between a public collection and the benchmark, and
they escaped in the same way: no aggregate figure depends on what a label means.

\subsection{A silent component failure}\label{sec:missed-tokenizer}

The arbitration tier ran through the entire recorded evaluation with a tokenizer
that inserted unknown-token identifiers between words, on a model that classifies
its own validation set perfectly. The class of fault is well documented:
inconsistent tokenization between training and inference degrades performance and
induces hallucination \cite{sun2023}, tokenization inconsistency undermines
robustness \cite{yan2025},
and tokens a model never saw in training produce anomalous behavior
\cite{land2024};
so is the systems-level consequence, in that entangled components make improving
one capable of worsening the whole \cite{sculley2015}, and optimizing each stage
of a cascade against its own local objective does not optimize the cascade
\cite{saberian2014}.

What no aggregate metric could show is that this had happened. Accuracy,
precision, recall, F1 and MCC are all reproduced exactly by a pipeline whose
neural tier is fed corrupted input, and the reason is structural rather than
statistical: routed a nearly pure population and permitted to escalate but not to
clear, the tier could not move the metric whatever it said. A component can
therefore fail completely without any end-to-end number shifting, provided the
architecture has already settled the outcome. The corollary is that metric
stability across versions, which is ordinarily read as evidence of a healthy
pipeline, carries no such information for a stage in this position.

A substitution run detects the condition in one pass. If replacing a component's
output with a constant leaves the metrics where they were, the component is not
contributing, and no amount of aggregate reporting will say so.

\subsection{A contamination episode found by audit}\label{sec:missed-contamination}

The second fault had the same signature. The project's history records 413
entries copied from the evaluation corpus into the similarity memory, and an
intermediate benchmark reporting $86.89\%$ recall while they were present. The
entries were removed before the canonical run, and we verify that the recorded
result is unaffected (Section~\ref{sec:results-corpus}).

The relevant point is not the magnitude, which is modest and confounded with a
concurrent configuration change, but the detection path. Nothing in the reported
metrics distinguished the contaminated run from a legitimate improvement, since
recall simply rose; the contamination came to light because someone deliberately
checked whether memory entries appeared in the evaluation corpus. An improvement
and a leak are indistinguishable at the level of a headline number, and only the
targeted check separates them. The same asymmetry has been reported at scale in
reproducibility studies of security machine learning
\cite{olszewski2023,schloegel2024}, which is why what follows is framed as a set
of reporting requirements rather than as advice.

\subsection{What to report}\label{sec:missed-recommendations}

Nine items follow, each answerable from artifacts a layered defense already
produces. The first four concern how a routed stage is described, the next two
are cheap correctness checks on the evaluation itself, the seventh concerns the
operating point, and the last two concern what the reported figure is compared
with and what its labels mean.

\begin{enumerate}
\item \textbf{The arbitration band base rate.} For every stage receiving routed
rather than complete traffic, report the class balance of the band it receives.
Without it an ablation figure for that stage cannot be interpreted
(Section~\ref{sec:baserate}).

\item \textbf{A substitution beside every deletion.} Report what happens when the
component's output is pinned to a constant, not only what happens when the
component is removed. The two answer different questions and, here, differ
4.8-fold.

\item \textbf{A transplant where a stronger component exists.} Put an
independently validated component in the stage and report how far the system
moves. Here a classifier separating the band at an AUC of $0.916$, against
$0.368$ for the original and $0.789$ for the repaired one, moved 16 decisions
(Section~\ref{sec:results-transplant}).

\item \textbf{The enforcement mapping.} State which internal states count as
positive. A pipeline converting an unresolved review into a block attributes
those blocks to whichever stage produced the review.

\item \textbf{Training--inference preprocessing parity.} Assert as a test that
the inference tokenizer reproduces the training tokenizer on a fixed probe set.
This cheap check would have caught the fault reported here.

\item \textbf{Per-source performance.} Report detection separately for each
corpus source. An aggregate over a corpus that is $82\%$ synthetic is dominated
by the generator rather than by the threat model.

\item \textbf{A held-out split for operating-point selection.} Select thresholds
on one partition and report on another. Where this is not done, say so.
For \sysname{} we measured the effect afterwards and found it small
(Section~\ref{sec:results-opt}); the point of the requirement is that this could
not have been known without the split.

\item \textbf{External baselines on the same records.} Run at least one public
detector over the identical records, with its operating point chosen away from
the reported data, and report it per source. Here it reverses the conclusion a
reader would draw from the aggregate (Section~\ref{sec:results-external}).

\item \textbf{Label provenance.} For every public collection inside a benchmark,
state how each record was derived from the original and what its label means
there. Three of the four collections here were altered on the way in, two of
them (BIPIA and WildJailbreak) in what their labels mean, and the fourth keeps its
texts but announces its labels through field names (Section~\ref{sec:results-labels}).
\end{enumerate}

\section{Limitations}\label{sec:limitations}

\paragraph{Scope} The study rests on one system and one corpus, with an
independent re-execution of the control on a second machine and an external
check set drawn from the original public collections. The
base-rate argument of Section~\ref{sec:baserate} is general, but its
instantiation here is a single measurement, and whether arbitration bands in
other layered defenses are as concentrated is an empirical question this paper
answers for one further system only. %
The two other defenses we measured do not route, so they cannot answer it; our
earlier gateway does, and the band its review model is consulted on is
$92.94\%$ malicious (Section~\ref{sec:results-cascade}).

\paragraph{One system, evaluated by its authors} Beyond the scope above, the
system was built in our own group. That is what made the switches, the repair
and the transplants possible, but it also means that we chose which interventions
to run and which findings to report. We have tried to limit this by reporting
every run we executed, including those that cut against the argument, and by
releasing the traces; a test on an independent routed system built by others
remains the most direct way to check the generality of the argument, and we have
not run one.

\paragraph{Enforcement mapping} The deletion--substitution
discrepancy is measured in the decisions the gateway enforces, with
\texttt{allow\_limited} counted as an allow, as the gateway itself counts it.
With that state enforced as a block the arbitrator's interventions change no
decision apart from 12 allow-hint releases (Section~\ref{sec:results-mapping}); the conclusion that its quality
does not reach the output survives, but the size of the discrepancy does not.

\paragraph{Measurement verified from traces, not re-executed} The released
bundle verifies every results table from per-request traces, but the source code
and the corpus texts belong to a graduate thesis in progress and are withheld, so
a reader cannot re-execute the runs, recompute the external detectors' scores on
the corpus, or repeat the corpus audit and the shortcut baselines without them.
The source code will be released once the thesis has been defended.

\paragraph{The repair intervention is opportunistic} It needs a component that is
genuinely defective and a defect that is genuinely mendable, and both conditions
happened to hold, which makes repair a diagnostic available once a fault is found
rather than a routine step.

\paragraph{Operating-point effect measured only over recorded points} The
configuration was selected on the evaluation corpus
(Section~\ref{sec:results-opt}); the held-out re-selection covers the eight
operating points that were actually run, not a continuous search over every
threshold the pipeline exposes.

\paragraph{Post-allow escalation not transplanted} Inputs that the
policy stage clears and a later enforcement rule then escalates account for 1,146
true positives and 75 false positives in the control run, and switching the step
off removes exactly these (Section~\ref{sec:results-postallow}). We isolated it by
deletion, which for a step of this kind removes no routing rule along with it,
and substituted its output with constant and random escalation, computed from the
traces and, for the constant, also executed; but we did not put a different
escalation rule in its place.

\paragraph{Residual non-determinism} One prompt of 25,000 received different
final decisions in two runs differing only in the tokenizer fix. The difference
originates upstream of the policy stage, in the similarity memory, whose
best-match score for that prompt takes one of two values across runs ($0.366$ or
$0.569$), so that the prompt registers a memory hit and bypasses the arbitrator
in some runs and is routed to it in others; the final decision changes only in
the pair where the arbitrator's verdict is pinned to ``benign''. It affects no reported
figure beyond the last digit; three repetitions of the control on one machine are
otherwise identical to the last stored digit.
On the second platform two executions of the control, one from each code state of
the transplant runs, agree on every final decision but differ in 16 semantic
scores, by at most $0.004$, and in the memory score of this prompt.

\paragraph{Shared mutable state} During one run, a separate service sharing the
same similarity-memory file wrote to it. That run's decisions are unaffected,
since all 97 distinct memory entries matched during it belong to the pinned
118-entry snapshot loaded at start and the pipeline does not reload the file at
runtime. We report it because the measurement protocol was not isolated from
concurrent writers, and a protocol relying on restored pinned state should be.

\paragraph{Memory figures withheld} Peak resident memory was recorded for every
run but is not reported: it did not vary consistently with configuration, and the
run making no arbitrator calls at all recorded a higher peak than the control, so
the measurement captured process-level noise rather than configuration cost.

\paragraph{An open question about rule provenance} The rescue branches match
formatting strings emitted by the corpus generator, yet were committed one to
fourteen days before that generator (and one to three weeks before the evaluation
corpus, Section~\ref{sec:results-decides}), so either the generator was written to produce
text the existing branches would catch, or the formatting convention predates
both. We report the coupling, which is measured, and not its direction, which is
not.

\section{Conclusion}\label{sec:conclusion}

A layered defense recorded $88.26\%$ accuracy with its neural arbitration tier
receiving corrupted input for the whole evaluation. Repairing that tier, which
lifts its corpus-wide AUC from $0.392$ to $0.855$, lowers the system's recall,
while replacing it with a constant answering ``malicious'' without reading its
input raises recall above both.

The explanation is not a defect but a structural property. Routing delivers the
arbitrator a band that is $96.82\%$ malicious, so a constant positive is right
$96.82\%$ of the time there; and the enforcement design lets a positive verdict
end in a block while a negative verdict releases an input only through the
\texttt{allow\_limited} branch, so classifier accuracy converts
into lost coverage rather than into avoided false positives. Under those two
conditions no arbitrator can gain more than the bound allows, and the deletion
ablation attributing 673 detections to it is measuring the change of status that
deletion imposes on the band instead.

We expect the reporting consequence to hold beyond the gateway studied here,
although we have measured it only there and in our earlier gateway. An
ablation figure for a routed stage is unreadable without the base rate of the
band that stage receives, and a deletion is not a substitute for a substitution.
Both are cheap; neither is standard.
Our earlier gateway, measured from its released records, shows the same
structure: the band its review model is consulted on is $92.94\%$ malicious, and a constant verdict differs
from its review model in 136 decisions when only blocks count as detections,
split almost evenly between them.

The same audit, extended past the components to the corpus and to the comparison,
moves the conclusion one step further. The largest single contributor to the
recorded detections is an enforcement step that appears in no architecture
description; the public collections inside the benchmark do not all mean what
their names say; and once operating points are chosen away from the test data, a
public classifier calibrated on the gateway's own synthetic records outperforms
the gateway on the public records taken together, in detection and in false
alarms at once. None of this is visible from the
headline figure, and all of it was visible from the records.

Transplanting that classifier into the gateway shows the same limit from the
other side. In the arbitration stage a classifier with an AUC of $0.916$ on the
band changes 16 decisions, about as many as a constant; in the semantic stage it
changes a thousand, and the gateway around it still detects fewer independent
attacks than the classifier does alone. What a stage contributes is fixed by its
position before its quality is considered, which is the case for reporting the
position with the result.

\section*{Data and Code Availability}

The corpus and model digests, per-request traces for all fifteen
runs, a verification script and
its output, and the pinned state snapshots are released as a reproduction bundle
\cite{barbican2026}, except two snapshots that store corpus texts, the similarity
memory and the semantic reference set, which are released by digest.
The source revision of the pipeline, the commits introducing the intervention
switches and the transplant hooks, the recorded diffs, the corpus texts and the
two text-bearing snapshots belong to a graduate thesis in progress and are not
released; as for our earlier gateway \cite{cascade}, the bundle serves the
verification of the reported figures rather than the re-execution of the system.
The source code is available to the editor and reviewers in confidence on request,
and will be released publicly once the thesis has been defended.
Every results table can be rebuilt from the released traces without them; the
divergence table and the shortcut baselines, which need the source and the
corpus texts, cannot.
Scripts are released under the MIT license and data and documentation under
CC~BY~4.0; third-party records keep their own licenses, and records derived from
BIPIA, together with the check set's WildJailbreak records, are distributed as
digests, against which their texts can be recovered from the original releases;
the check set's LLMail-Inject texts and the InjecAgent records in their own
format are included.
The bundle further carries the traces of the second series, of the
second-platform control and of the transplant, LLMail line-removal,
header-removal, substitution and InjecAgent runs, the per-component outputs of LLM Guard and NeMo Guardrails, the
external detectors' per-record scores with model
revisions, the lineage-group assignment, the provenance checks of the public
records with the per-record WildJailbreak type mapping, the scripts that recompute
every table from these files, and a script that asserts the figures of the main
tables against the traces.

\printcredits

\section*{Funding}

This research did not receive any specific grant from funding agencies in the
public, commercial, or not-for-profit sectors.

\section*{Declaration of Competing Interests}

The authors declare no known competing financial interests or personal
relationships that could have appeared to influence the work reported in this
paper. They disclose one non-financial interest: \sysname{} was developed by the
authors' own research group. The measurements
reported here were produced by executing a pinned revision of its source, and the
artifacts supporting the reported measurements are released with the paper.

\section*{Declaration of generative AI and AI-assisted technologies in the manuscript preparation process}

During the preparation of this work the authors used Claude (Anthropic) for
literature screening, for drafting and editing prose, and for cross-checking
arithmetic against the released traces. No figure, table entry or measurement reported here
was produced by a language model: every number derives from executing the
systems described in Section~\ref{sec:method} and recomputing from the resulting
traces, or from the released records of our earlier gateway. The authors reviewed and edited all content and take full
responsibility for it.

\end{document}